\documentclass[11pt,a4paper]{article}
\usepackage{jheppub}

\usepackage{bm}
\usepackage{amsmath}
\usepackage{graphicx}
\usepackage{float}
\usepackage{setspace}
\usepackage{footmisc}
\usepackage[makeroom]{cancel}
\usepackage[utf8]{inputenc}
\usepackage{bbold}
\usepackage{comment}
\definecolor{darkgreen}{RGB}{0,120,0}

\newcommand{\prevchangeflag}[1]{{ #1}}

\newcommand{\changeflag}[1]{{ #1}}

\def\be{\begin{equation}}
\def\ee{\end{equation}}
\def\ba{\begin{eqnarray}}
\def\ea{\end{eqnarray}}

\usepackage{tabularx}

\usepackage{listings}
\usepackage{fancyvrb}
\usepackage{amsfonts}

\newcommand{\mbf}{\mathbf}
\newcommand{\mrm}{\mathrm}
\newcommand{\Tr}{\mrm{Tr}}

\newcommand{\ah}{_{\mathrm{ah}}}
\newcommand{\cl}{{\mathrm{cl}}}

\newcommand{\fig}{Fig.~}

\newcommand{\se}{Sec.~}

\newcommand{\re}{Ref.~}
\newcommand{\res}{Refs.~}
\newcommand{\app}{App.~}
\newcommand{\eq}{Eq.~}

\newcommand{\pToFigs}{.}

\begin{document}

\title{The imaginary part of the heavy-quark potential from real-time Yang-Mills dynamics}

\author[a]{Kirill Boguslavski}
\affiliation[a]{Institute for Theoretical Physics, Technische Universit\"at Wien, 1040 Vienna, Austria}

\author[b]{Babak S. Kasmaei}
\author[b]{Michael Strickland}
\affiliation[b]{Department of Physics, Kent State University, Kent, OH 44242, United States}

\emailAdd{mstrick6@kent.edu}
\emailAdd{bsalehik@kent.edu}
\emailAdd{kirill.boguslavski@tuwien.ac.at}

\abstract{
We extract the imaginary part of the heavy-quark potential using classical-statistical simulations of real-time Yang-Mills dynamics in classical thermal equilibrium.   
The $r$-dependence of the imaginary part of the potential is extracted by measuring the temporal decay of Wilson loops of spatial length $r$.
We compare our results to continuum expressions obtained using hard thermal loop theory and to semi-analytic lattice perturbation theory calculations using the hard classical loop formalism.  
We find that, when plotted as a function of $m_D r$, where $m_D$ is the hard classical loop Debye mass, the imaginary part of the heavy-quark potential shows little sensitivity to the lattice spacing at small $m_D r \lesssim 1$ and agrees well with the semi-analytic hard classical loop result.  
For large quark-antiquark separations, we quantify the magnitude of the non-perturbative long-range corrections to the imaginary part of the heavy-quark potential.  We present our results for a wide range of temperatures, lattice spacings, and lattice volumes. 
This work sets the stage for extracting the imaginary part of the heavy-quark potential in an expanding non-equilibrium Yang Mills plasma.
}

\arxivnumber{2102.12587}

\maketitle
\flushbottom

%%%%%%%%%%%%%%%%%%%%%%%%%%%%%%%%%%%%%%%%%%%%%%%%%%%%%%%%%%%%%
\section{Introduction}
%%%%%%%%%%%%%%%%%%%%%%%%%%%%%%%%%%%%%%%%%%%%%%%%%%%%%%%%%%%%%

At extreme energy densities nuclear matter undergoes a phase transition from a state characterized by confined hadrons to a state in which quarks and gluons become the relevant degrees of freedom.   Ultrarelativistic heavy-ion collision experiments at the Relativistic Heavy Ion Collider in New York and the Large Hadron Collider at CERN have now collected a wealth of data concerning the properties of the high energy density phase of nuclear matter, which is called the quark-gluon plasma (QGP) \cite{Averbeck:2015jja}.  One of the key observables used in measuring the properties of the QGP is the suppression of heavy quark-antiquark bound states such as bottomonium.  The suppression observed by experimentalists gives a measurement of the in-medium breakup rate of bottomonium states and allows one to constrain key plasma observables such as the initial central temperature of the QGP generated in heavy-ion collisions, its shear viscosity to entropy density ratio, and the differential rate at which the QGP expands in non-central collisions (see e.g. \cite{Acharya:2020kls,ATLAS5TeV,Sirunyan:2018nsz,Ye:2017vuw}).  

Fundamentally, the computation of the survival probability of a given bottomonium state can be cast into the framework of open quantum systems (OQS) in which there is a probe (bottomonium states) and medium (light quarks and gluons).  Within the OQS framework, in order to describe the in-medium evolution of bottomonium states one must trace over the medium degrees of freedom and obtain evolution equations for the reduced density matrix of the system \cite{Akamatsu:2011se,Akamatsu:2012vt,Akamatsu:2014qsa,Katz:2015qja,Brambilla:2016wgg,Kajimoto:2017rel,Brambilla:2017zei,Blaizot:2017ypk,Blaizot:2018oev,Yao:2018nmy,Alund:2020ctu,Islam:2020gdv,Islam:2020bnp,Brambilla:2020qwo}.  In the limit that the medium relaxation time scale and the intrinsic time scale of the probe are much smaller than the probe relaxation time scale, the resulting dynamical equation for the reduced density matrix can be cast into a so-called Lindblad form \cite{lindblad,gorini}.  A key outcome of such calculations is that the effective heavy-quark potential possesses an imaginary part which can be related to the total in-medium decay width of the states.  This imaginary part has been determined using direct quantum field theoretic or effective field theory calculations \cite{Laine:2006ns,Brambilla:2016wgg,Brambilla:2017zei}.

There have been computations of the imaginary part of the heavy-quark potential based on high-temperature quantum chromodynamics (QCD) calculations in the hard thermal loop (HTL) limit \cite{Laine:2006ns,Dumitru:2007hy,Brambilla:2008cx,Burnier:2009yu,Dumitru:2009fy,Dumitru:2009ni,Margotta:2011ta,Guo:2018vwy}, using effective field theory (pNRQCD) \cite{PhysRevD.21.203,Lucha:1991vn,Brambilla:2004jw,Brambilla:2010xn}, finite-temperature lattice QCD \cite{Rothkopf:2009pk,Rothkopf:2011db,Burnier:2012az,Burnier:2013nla,Burnier:2015nsa,Burnier:2015tda,Burnier:2016mxc,Burnier:2016kqm,Bala:2019cqu,Bala:2020tdt}, and real-time classical-statistical solutions of Yang-Mills theory in classical thermal equilibrium
\cite{Laine:2007qy,Lehmann:2020fjt}.  In this work we build upon the studies presented in Ref.~\cite{Laine:2007qy} and present findings that are complementary to Ref.~\cite{Lehmann:2020fjt}. In Ref.~\cite{Laine:2007qy} the authors presented first results for the imaginary part of the heavy-quark potential using classical-statistical Yang-Mills simulations on spatially 3D lattices of size $12^3$ and $16^3$.  In this paper we extend these results to larger lattices up to $252^3$
and consider SU(2) and SU(3) gauge theories.  Additionally, the results of Ref.~\cite{Laine:2007qy} were presented only for a few spatial points $r$ in a table. Due to the use of rather large lattice sizes, we can now compute the imaginary part of the heavy quark potential at larger values of $r/a$ and reconstruct the functional form of the imaginary part of the heavy-quark potential for a much wider range of distances.
This allows us to make more precise comparisons between our lattice-extracted imaginary part and (a) results obtained in the continuum limit using hard thermal loops \cite{Laine:2006ns} and (b) results obtained using the lattice-regularized hard-classical-loop (HCL) theory \cite{Laine:2007qy}.

Herein we will present results for the imaginary part of the heavy-quark potential obtained using classical Yang Mills (CYM) simulations of a thermalized gluonic plasma.  The use of CYM simulations is motivated by the fact that in situations where (a) gluonic occupation numbers are large, such as in thermal equilibrium for sufficiently low momenta or in the initial stages of heavy-ion collisions, and (b) the gauge coupling is weak $g^2 \ll 1$, vacuum contributions to observables are suppressed by powers of the gauge coupling. As a consequence, neglecting vacuum contributions in such cases often leads to a good approximation to the full quantum dynamics. \changeflag{However, one should note that in the real-time classical lattice theory rotational invariance is broken and the short and long distance physics do not decouple \cite{Bodeker:1995pp, Arnold:1997yb}. As a consequence, the results that we obtain in our classical simulations may differ from those in thermal equilibrium.}

In order to extract information about thermal systems using CYM, one must prepare the thermalized field configurations.  Historically, this is done by preparing 3+1D configurations using Monte-Carlo techniques \cite{Grigoriev:1988bd,Ambjorn:1990pu,Ambjorn:1997jz,Moore:1996qs,Moore:1999fs,Hart:2000ha}.  Herein we follow a simpler strategy and initialize the system close to thermal equilibrium using momentum-space initialization and let the fields thermalize in real time before extracting observables. 
The advantage of this procedure is that we can run simulations on very large lattices with small lattice spacings $a$ at moderate computational cost. In practice, however, this means that we have to extract the temperature $g^2 T$ numerically in the thermalized system from correlation functions. We find that this only introduces a small uncertainty for $g^2 T$ of the order of a few percent, which justifies this approach. 

One additional complication that arises when dealing with CYM treatments is that they do not possess a finite ultraviolet limit due to the Rayleigh-Jeans divergence \changeflag{and the theory is non-renormalizable} \cite{Kajantie:1993ag,Ambjorn:1995xm,Arnold:1996dy,Moore:1999fs,Berges:2013lsa,Epelbaum:2014yja}.  For this reason, it is important to identify a suitable manner in which to scale results in order to extract relevant information. We will use the Debye mass $m_D$ computed in the hard-classical-loops framework to demonstrate that, when plotted as a function of $m_D r$, the imaginary part of the heavy-quark potential \changeflag{is only mildly sensitive to the lattice spacing, or more generally to the simulation parameter $\beta \propto 1 / (g^2Ta)$, at small distances $m_D r \lesssim 1$.}
Additionally we find that, when presented in this manner, results obtained using CYM simulations agree well with the semi-analytic hard classical loop result at small quark-antiquark separations.  
Our study also suggests that the latter approaches a finite form in the large-$\beta$ limit for all separations.
We present our results for a wide range of temperatures, lattice spacings, and lattice volumes, which map to values of $\beta$ in the range $16 \rightarrow 300$.  

This work sets the stage for extracting the out-of-equilibrium imaginary part of the heavy-quark potential in expanding Yang Mills plasmas \cite{Berges:2013eia,Berges:2013fga}. 
\changeflag{Classical Yang-Mills simulations are applicable in that case, allowing one to extrapolate to small lattice spacings, because the dynamics of the highly occupied plasmas are governed by hard scales that can be set below the lattice momentum cutoff.} 
Such non-equilibrium extractions are necessary since it is currently unknown how to analytically extract the heavy-quark potential in anisotropic plasmas due the presence of a non-Abelian plasma instability called the chromo-Weibel instability \cite{Mrowczynski:1996vh,Romatschke:2003ms,Arnold:2003rq,Burnier:2009yu,Nopoush:2017zbu}. 
In particular, recent studies have shown that particularly in highly anisotropic systems nonperturbative effects beyond hard loop calculations are crucial \cite{Boguslavski:2019fsb,Boguslavski:2021buh}. 

The structure of our paper is as follows.  In Sec.~\ref{sec:theory} we provide the theoretical background necessary for the computation of the imaginary part of the heavy-quark potential.  In Sec.~\ref{sec:results} we present the results from lattice simulations and the comparison \changeflag{to perturbative calculations.} We present concluding remarks and outlook for future works in Sec.~\ref{sec:conclusions}. 

%%%%%%%%%%%%%%%%%%%%%%%%%%%%%%%%%%%%%%%%%%%%%%%%%%%%%%%%%%%%%
\section{Theory and numerical setup}
\label{sec:theory}
%%%%%%%%%%%%%%%%%%%%%%%%%%%%%%%%%%%%%%%%%%%%%%%%%%%%%%%%%%%%%

We consider pure SU($N_c$) gauge theory with the Yang-Mills classical action
\begin{align}
 S[A] = -\frac{1}{2}\int d^4 x\, \Tr \left(F^{\mu\nu} F_{\mu\nu}\right),
\end{align}
with Einstein sum convention for repeated Lorentz indices $\mu,\nu = 0,\dots,3$. 
The field strength tensor is given by $F_{\mu\nu} = \partial_\mu A_\nu - \partial_\nu A_\mu -ig [A_\mu,A_\nu]$ with gauge coupling $g$, gauge field $A_\mu(x)$ and commutator $[,]$. Unless stated otherwise, we will use $N_c=3$.

\subsection{Lattice discretization and equations of motion}

We use a standard real-time lattice discretization approach where fields are discretized on cubic lattices with $N^3$ sites and lattice spacing $a$ (see, e.g., \res\cite{Boguslavski:2018beu,Berges:2013fga} and references therein for more details). In this real-time approach, spatial gauge fields are replaced by gauge links $U_j(t,\mbf x) \approx \exp\left( i g a A_j(t,\mbf x) \right)$ at discrete coordinates $x_k = n_k a$ for $n_k = 0, \ldots, N-1$, while temporal gauge with $A_0 = 0$, and thus $U_0 = \mathbb{1}$ is used. The classical equations of motion are written in a gauge-covariant manner as
\begin{align}
\label{eq:classEOM}
 U_j(t+a_t/2,\mbf x) &= \exp\left( i a_t a g E^j(t,\mbf x) \right)U_j(t-a_t/2,\mbf x) \, , \nonumber \\
 gE^j(t+a_t,\mbf x) &= gE^j(t,\mbf x) 
 - \frac{a_t}{a^3} \sum_{j \neq i} \left[ U_{ij}(t-a_t/2,\mbf x) + U_{i(-j)}(t-a_t/2,\mbf x) \right]\ah \, ,
\end{align}
and are solved alternately in a leapfrog scheme~\cite{Krasnitz:1995xi}. The plaquette is defined as $U_{ij}(\mbf x) = U_i(\mbf x)U_j(\mbf x + \mbf{a}_i)U_i^\dagger(\mbf x + \mbf{a}_j)U_i^\dagger(\mbf x)$, and analogously for $U_{i(-j)}$, with vectors $\mbf{a}_{i/j}$ of length $a$ in spatial 
directions $i,j = 1,2,3$. $[V]\ah \equiv -i(V-V^\dagger - \Tr(V-V^\dagger)/N_c)/2$ defines the anti-Hermitian traceless part of the matrix $V$. The time step is taken to be small, typically $a_t/a = 0.01$ in this work, to reduce temporal lattice artifacts. The classical equations \eqref{eq:classEOM} guarantee that the degree of Gauss law violation is conserved at every time step
%
%\begin{widetext}
\begin{align}
 D_j E^j(t,\mbf x) &\equiv \frac{1}{a}\sum_j\left( E^j(t,\mbf x) - U_j^\dagger(t-a_t/2,\mbf x - \mbf{a}_j)E^j(t,\mbf x - \mbf{a}_j)U_j(t-a_t/2,\mbf x - \mbf{a}_j) \right) \nonumber \\
 &= \text{constant} \,.
\end{align}
%\end{widetext}
%
Moreover, this lattice discretization guarantees gauge invariance of observables like Wilson loops that are supposed to be gauge invariant.

\subsection{Computation of observables}

Classical observables are computed by averaging over different configurations evolved independently 
\begin{align}
 \langle O \rangle(t) \equiv \frac{1}{n}\sum_{j=1}^n O_j(t)\,,
\end{align}
and their uncertainity corresponds accordingly to the uncertainty of the mean.

\subsection{Imaginary part of the classical potential}
\label{sec:theory_potential}

We are interested in extracting the imaginary part of the classical potential $V_\cl(t,r)$ with $r \equiv |\mbf x|$. Following \res\cite{Laine:2006ns,Laine:2007qy}, it can be calculated using
\begin{align}
 i\partial_t C_\cl(t,r) = V_\cl(t,r) C_\cl(t,r) \, ,
\end{align}
as the asymptotic temporal slope of $\log[C_\cl(t,r)]$. The classical thermal Wilson loop  $C_\cl(t,r)$ is defined as
\begin{align}
 C_\cl(t,r) \equiv \frac{1}{N_c}\,\Tr \left\langle W[(t_0,\mbf x);(t,\mbf x)]\, W[(t,\mbf x);(t,\mbf 0)]\, W[(t,\mbf 0);(t_0,\mbf 0)]\, W[(t_0,\mbf 0);(t_0,\mbf x)] \right\rangle,
\end{align}
with temporal Wilson lines $W[(t_0,\mbf x);(t,\mbf x)] = \mathbb{1}$ and spatial Wilson lines $W[(t,\mbf 0);(t,\mbf x)] = U_j(t,\mbf 0)U_j(t,\mbf{a}_j)U_j(t,2\,\mbf{a}_j) \cdots U_j(t,\mbf x)$ for $\mbf x = \mbf{\hat a}_j\,r$ and $\mbf{\hat a}_j = \mbf{a}_j/a$ being a spatial unit vector. Since the classical thermal state is homogeneous, the Wilson loop is additionally averaged over all lattice points by averaging over the reference coordinates $\mbf 0$. 

In order to extract the imaginary part of $V_\cl(t,r)$ we compute the time-dependence of $C_\cl(t,r)$.  Due to the imaginary part of the in-medium heavy quark potential, this quantity will decay exponentially at late times with the rate of exponential decay set by the imaginary part of $V_\cl(t,r)$.  As a result, $V_\cl(t,r)$ can be computed using the logarithmic slope of this decay as mentioned above.  We note that this is formally applicable only in the large-$t$ limit, but we find that an exponential decay is established rather quickly, particularly when using large lattice sizes. We then define the imaginary part of the static classical potential as the late-time limit 
\begin{align}
 {\rm Im}[V_{\rm cl}(r)] \equiv \lim_{t \to \infty} {\rm Im}[V_{\rm cl}(t,r)]\,.
\end{align}
In practice, this agrees with the logarithmic slope extracted from $C_\cl(t,r)$.

\subsection{Construction of quasi-thermalized and fully-thermalized classical states}
\label{sec:thermalstate}

In order to associate our measurements of $V_\cl(t,r)$ with a specific temperature $T$, it is necessary to have a method for self-consistently determining the temperature of the CYM fields.  One can construct thermalized CYM configurations using 3+1D Monte-Carlo techniques \cite{Grigoriev:1988bd,Ambjorn:1990pu,Ambjorn:1997jz,Moore:1996qs,Moore:1999fs,Hart:2000ha}. However, as mentioned in the introduction, we use a simpler technique, which amounts to initializing the fields in a quasi-thermal configuration in momentum-space, as will be explained around \eqref{eq:EE_noGauss_init}, and then allowing them to self-thermalize dynamically.  On large lattices this method is quite efficient and one finds that the fields self-thermalize quickly.  One complication is that the resulting equilibrated temperature $T$ is different than the quasi-thermal initial temperature $T_0$ used to initialize the fields in momentum-space.  As a result, one must have a method to extract the time-dependent temperature in order to (a) determine when field thermalization is functionally complete and (b) extract the final thermalized temperature of the CYM fields.  Once the system is fully thermalized (within an acceptable uncertainty), one can then proceed with the measurement of $V_\cl(t,r)$.

For this purpose, we extract chromo-electric field correlation functions in Fourier space. In practice, the chromo-electric fields are Fourier transformed as 
\begin{align}
E^j(t,\mbf p) = a^3\sum_{\mbf x} e^{-2\pi i \sum_k n_k m_k/N} E^j(t,\mbf x) \, , 
\end{align}
and projected onto normalized transverse vectors $v_j^\lambda(\mbf p)$ with two polarizations $\lambda = 1,2$ transverse to the longitudinal polarization vector $v_j^{3}(\mbf p) = p_j/p$,
providing $E_\lambda(t,\mbf p)$. Here we used $x_k = n_k\, a$ with $k=1,2,3$, the lattice momentum definition $p_k\, a = -i(1-
\exp(-2\pi i m_k/N))$, $p^2 = \sum_k |p_k|^2$, and $n_k, m_k = 1,\dots, N$. The correlation functions are then computed as
\begin{align}
 \langle EE \rangle_{T}(t, p) &= \frac{1}{2(Na)^3(N_c^2-1)}\sum_{\lambda = 1,2} \left\langle |E_{\lambda}(t,\mbf p)|^2 \right\rangle \nonumber \\
 \langle EE \rangle_{L}(t, p) &= \frac{1}{(Na)^3(N_c^2-1)} \left\langle |E_{\lambda=3}(t,\mbf p)|^2 \right\rangle\,,
\end{align}
and similarly for $A$ fields. This expression implies averaging over the direction of $\mbf p$ due to the system's approximate isotropy for not too large momenta and we will consider $\langle EE \rangle_{T/L}$ as functions of $p$.\footnote{
Although the anisotropy of the cubic lattice becomes more prominent close to the lattice cutoff at $|p_k| = 2/a$, we will mostly use these correlators as means to extract parameters by fitting to them over a large range of momenta, such that the anisotropic behavior of the cube does not play a significant role here.
}
Note that these correlation functions are, in general, gauge dependent. We get rid of the residual gauge freedom by gauge-fixing to the Coulomb-type gauge $\partial^j A_j(t,\mbf x) = 0$ to almost machine precision right before computing the correlators.

As mentioned above, while all measurements will be performed in classical thermal equilibrium, for practical purposes we start our simulations in a state very close to thermal equilibrium and let the system dynamically thermalize. We use two-point correlation functions $\langle EE \rangle$ in Fourier space to check to what extent the system has thermalized. Our procedure is as follows.

We set the fields in Fourier space at initial time $t = 0$ such that they satisfy
\begin{align}
\label{eq:EE_noGauss_init}
 g^2\langle EE \rangle_T = g^2T_0\,, \qquad g^2\langle EE \rangle_L = 0\,, \nonumber \\  g^2\langle AA \rangle_T = \frac{g^2T_0}{p^2}\,, \qquad g^2\langle AA \rangle_L = 0\,.
\end{align}
These initial conditions can be implemented efficiently by setting (see, e.g., \cite{Smit:2002yg,Berges:2013fga,Boguslavski:2018beu})
\begin{align}
\label{eq:AE_initial}
 gE^j_a(t{=}0,\mbf p) &= \sqrt{g^2T_0}\,\sum_{\lambda = 1,2} v_j^{\lambda}(\mbf p)\,\beta_a^{\lambda}(\mbf p)  \nonumber \\
 gA_j^a(t{=}0,\mbf p) &= \sqrt{\frac{g^2 T_0}{p^2}}\,\sum_{\lambda = 1,2} v_j^{\lambda}(\mbf p)\,\alpha_a^{\lambda}(\mbf p)\,,
\end{align}
with the adjoint color index $a = 1, \dots, N_c^2-1$, and with Gaussian distributed complex random numbers satisfying $\langle (\alpha_a^{\lambda}(\mbf p))^* \alpha_b^{\lambda'}(\mbf p') \rangle = \delta_{ab}\delta_{\lambda \lambda'}\delta_{\mbf p \mbf p'} N^3a^3$, $\langle (\beta_a^{\lambda}(\mbf p))^* \beta_b^{\lambda'}(\mbf p') \rangle = \delta_{ab}\delta_{\lambda \lambda'}\delta_{\mbf p \mbf p'} N^3a^3$, while all other correlations vanish.
Subsequently, we restore the Gauss law $D_j E^j(x) = 0$ with covariant derivative $D_\mu$ to machine precision using the algorithm in \re\cite{Moore:1996qs}. This algorithm changes the chromo-electric field correlators mostly at low momenta
and, for instance, makes $\langle EE \rangle_L$ finite. Then we let the system evolve according to the Yang-Mills classical equations of motion \eqref{eq:classEOM}. Classical thermal equilibrium is reached when the $\langle EE \rangle_{T/L}(t,p)$ correlation functions do not visibly change over an extended amount of time.

%%%%%%====FIGURE======%%%%%%%%%%%%%%%%%%%%%%%%%%%%%
\begin{figure}[t!]
\centerline{
\includegraphics[width=0.48\linewidth]{\pToFigs/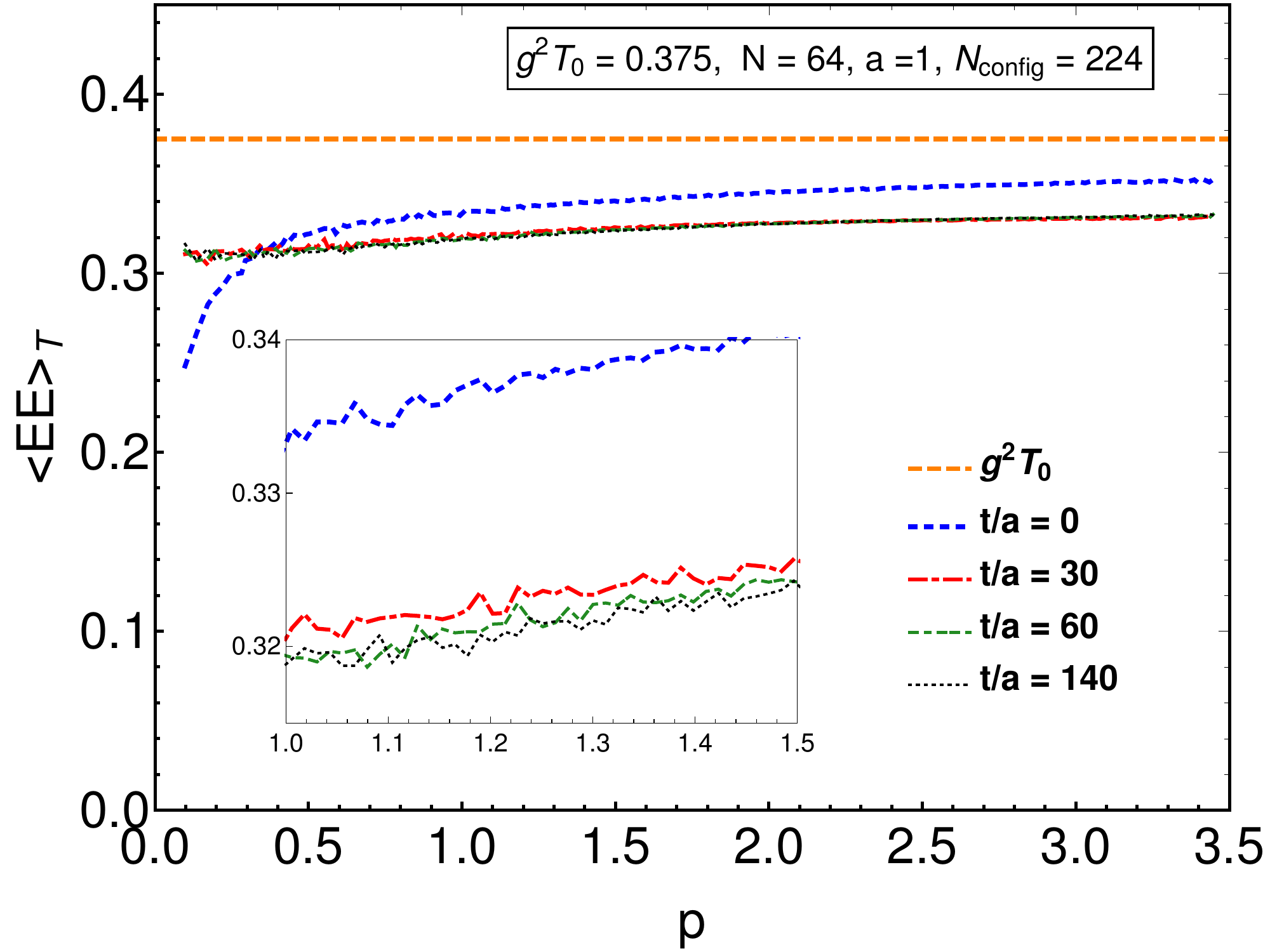}
$\;\;\;$
\includegraphics[width=0.48\linewidth]{\pToFigs/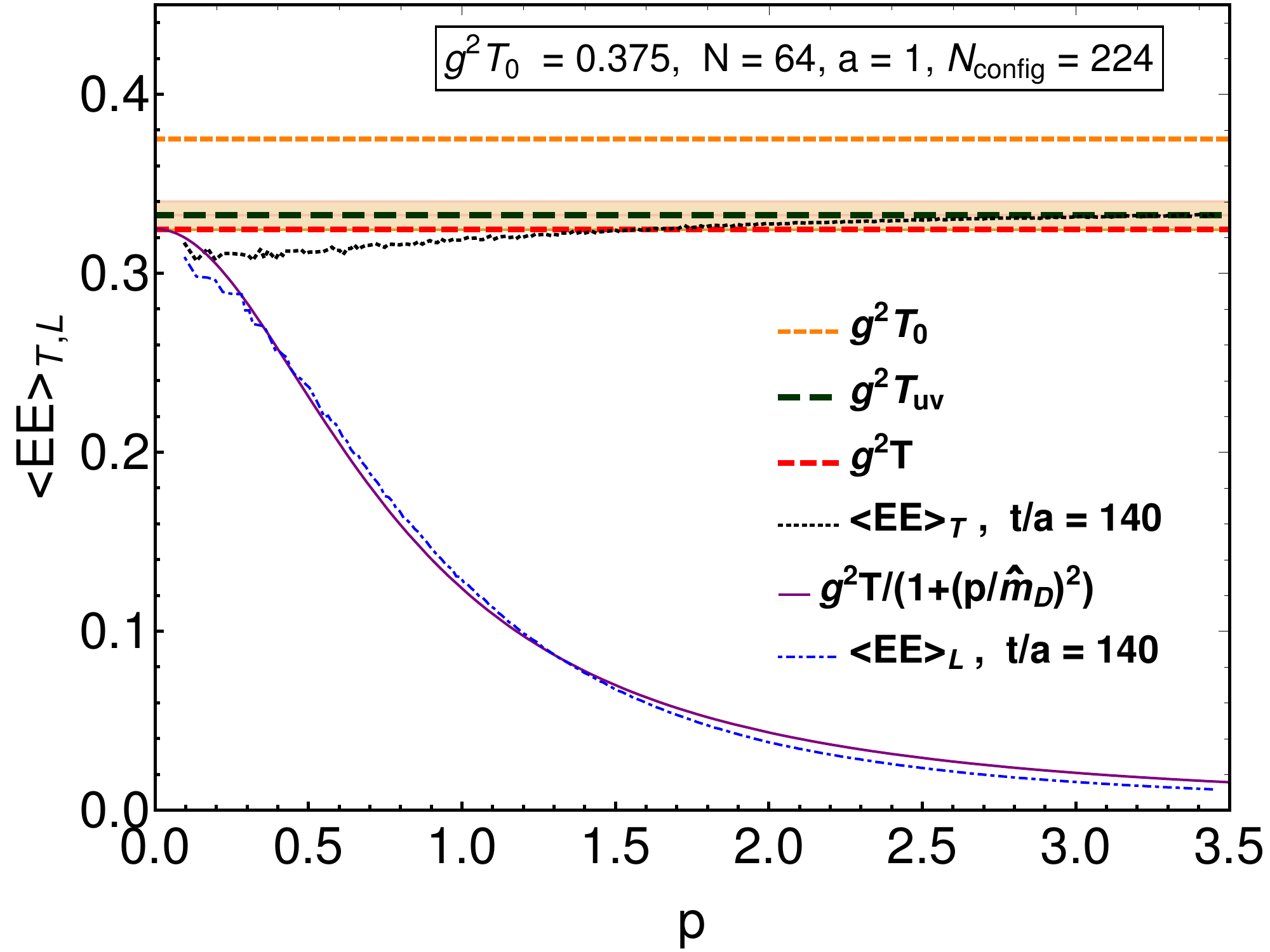}
}
\caption{
(Left:) Example of thermalization for $g^2 T_0 = 3/8$ and $a=1$. The transverse chromo-electric field correlation is shown at different times of the simulated gauge dynamics. 
The inset shows a zoomed region of the plot. The dashed orange line corresponds to the field initialization \eqref{eq:EE_noGauss_init} with $g^2 T_0$ before the Gauss law is restored.
(Right:) Extracting $T$ and $\hat{m}_D$ by fitting \eqref{eq:extractT} to transverse and longitudinal chromo-electric field correlations. The fit functions are given by the black dashed and continuous purple lines. The red dashed line corresponds to the value $g^2 T_{\text{UV}}$ at the hardest momentum, the shaded error band shows the systematic error in the estimation of the temperature $[T-\Delta T, T+\Delta T]$ with $\Delta T = T_{\text{UV}} - T$ . 
} 
\label{plot:T_md_convergence}
\end{figure}
%%%%%%%%%%%%%%%%%%%%%%%%%%%%%%%%%%%%%%%%%%%%%%%%%%%%

An example of the results obtained using this thermalization algorithm is shown in the left panel of \fig\ref{plot:T_md_convergence} for $\langle EE \rangle_{T}$ and simulation parameters $g^2 T_0 = 0.375$, $N=64$, $a=1$, averaged over $N_{\text{config}} = 224$ configurations. The dashed orange horizontal line shows the initial correlator prior to Gauss law restoration, that equals $g^2 T_0$ for all momenta. After the restoration, the transverse correlator shows some momentum dependence and is given by the blue dashed line at $t/a = 0$. In the subsequent evolution, the system thermalizes rather quickly. Already for times $t/a \gtrsim 30$, the correlation function is almost stationary, for later times $t/a \gtrsim 60$, the deviation between curves at different times lies within the uncertainty of the data, as can be seen in the inset. At that point, the system can be considered approximately thermal.

As shown in \re\cite{KM2012} and recently used in \cite{Matsuda:2020qwx}, the correlation functions $\langle EE \rangle_{T/L}(t,p)$ encode the temperature and the Debye mass of the classical state
\begin{align}
\label{eq:extractT}
 g^2\langle EE \rangle_T \approx g^2T\,, \qquad g^2\langle EE \rangle_L \approx g^2T\,\frac{\hat{m}_D^2}{p^2+\hat{m}_D^2}\,.
\end{align}
These relations are not exact but become more reliable at larger momenta $p \gg m_D$. We can use them to extract the temperature $g^2 T$ and an estimate for the Debye (screening) mass $\hat{m}_D$. 
A systematic error (for $g^2 T$) is estimated as the deviation of the correlators from these simple functional forms, as will be explained shortly. 
The extraction of these parameters 
from the transverse and longitudinal chromo-electric field correlations is demonstrated in the right panel of \fig\ref{plot:T_md_convergence} for the same parameters as in the left panel at the time $t/a = 140$. One can see that the thermalized correlations are lower than the initial value $g^2 T_0$ for the considered parameters and that they indeed agree well with the fit functions $g^2\langle EE \rangle_T^{\text{fit}} = g^2\hat{T}_{E_tE_t}$ and $g^2\langle EE \rangle_L^{\text{fit}} = g^2\hat{T}_{E_lE_l}/(1+p^2/\hat{m}^2_D)$, as shown by the black dashed and purple continuous lines. Remarkably, these independent fits lead to almost the same value of the temperature and we set
\begin{align}
 g^2T = g^2\hat{T}_{E_tE_t}\,.
\end{align}
Since the transverse correlator $\langle EE \rangle_T$ still grows slowly with $p$, we extract the value $g^2 T_{\text{UV}}$ at the largest momentum to estimate the error as $\Delta T = T_{\text{UV}} - T$. This systematic error is shown by the shaded band in the figure. We find that the uncertainty in the extraction of the temperature is generally quite low and becomes smaller with decreasing lattice spacing. For the coarse lattice spacing $a=1$ used in \fig\ref{plot:T_md_convergence}, the relative error is only $\Delta T/T < 3\%$. 

%%%%%%====FIGURE======%%%%%%%%%%%%%%%%%%%%%%%%%%%%%
\begin{figure}[t]
\centerline{
\includegraphics[width=0.48\linewidth]{\pToFigs/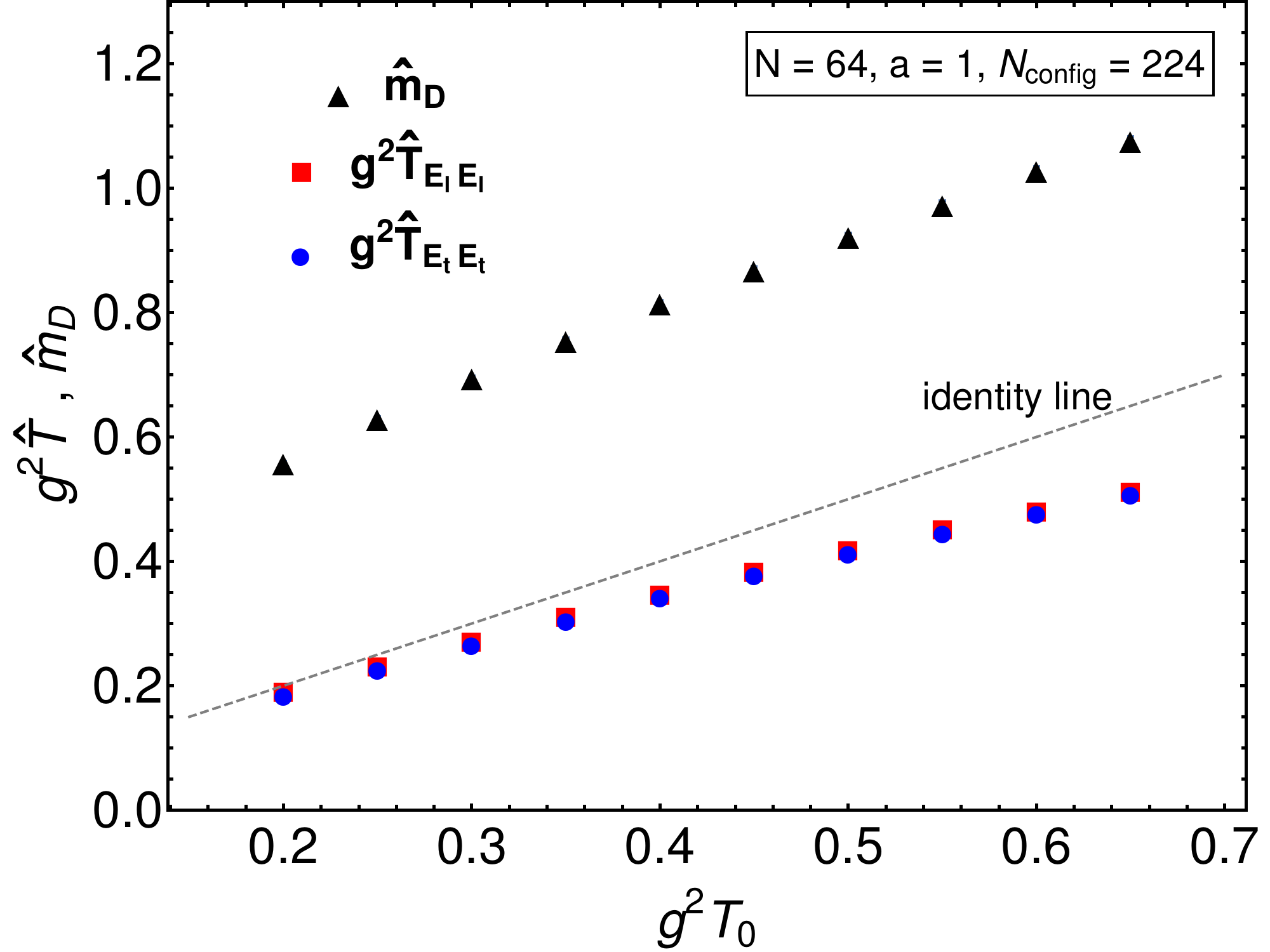}
$\;\;\;$
\includegraphics[width=0.48\linewidth]{\pToFigs/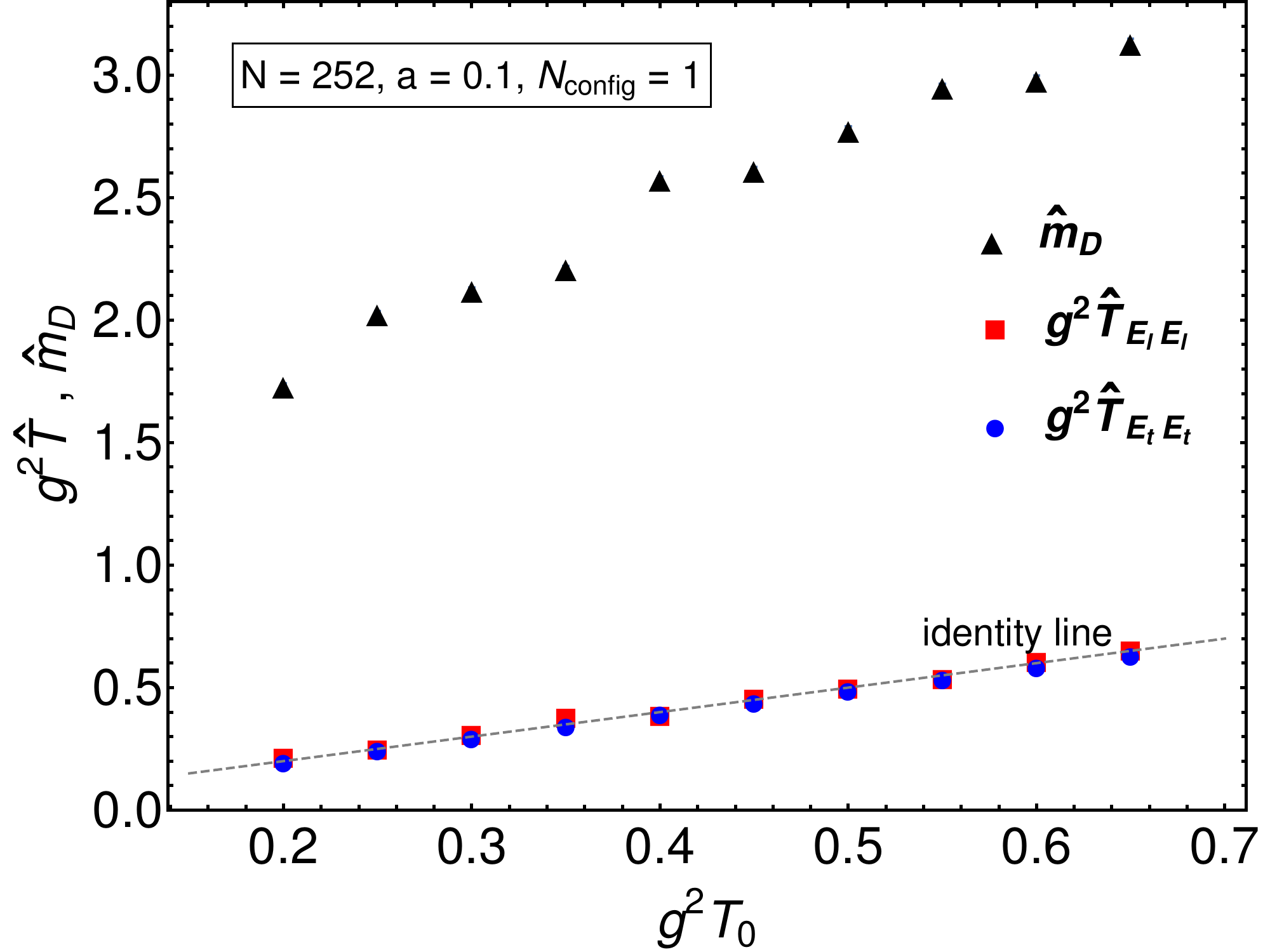}
}
\caption{Extracted values of $g^2\hat T_{E_tE_t}$, $g^2\hat T_{E_lE_l}$ and $\hat{m}_D$ for different values of $g^2 T_0$ for $a = 1$ (left) and $a = 0.1$ (right). All parameters are shown in lattice units.} 
\label{plot:Tmd_T0}
\end{figure}
%%%%%%%%%%%%%%%%%%%%%%%%%%%%%%%%%%%%%%%%%%%%%%%%%%%%

The extracted temperature and mass values, $g^2\hat T_{E_tE_t}$, $g^2\hat T_{E_lE_l}$ and $\hat{m}_D$, are shown in \fig\ref{plot:Tmd_T0} as functions of $g^2 T_0$ for the larger lattice spacing $a = 1$ (left) and a fine lattice spacing $a = 0.1$ (right). One observes that both $g^2\hat T_{E_tE_t}$ and $g^2\hat T_{E_lE_l}$, visible as blue and red points, respectively, agree well in both cases. While for the small lattice spacing, $g^2 T_0$ and the extracted $g^2 T$ are almost identical, one observes deviations that grow with temperature for the coarser lattice. Since the lattice simulations depend on the combination $\beta = 2N_c / (g^2Ta)$, as will be explained in \se\ref{sec:coupling_scaling}, we can summarize that deviations increase with decreasing $\beta$, while for growing $\beta$ values, both $\Delta T/T$ and $|T_0-T|/T$ become smaller, enabling a better control of the value of the temperature.

\subsection{The HTL, HCL, and lattice Debye masses}

We will also need to compute the Debye mass in order to compare results obtained using different analytic and numerical approaches. The definition of the Debye mass 
depends on the method being used.  
In this paper we consider three methods for defining the Debye mass: (1) continuum hard-thermal-loops calculations, (2) lattice-discretized hard classical loops calculations, and (3) direct lattice measurement using chromo-electric field correlators.  The definitions in these three cases are:

\begin{enumerate}
\item 
The well-known continuum hard thermal loop (HTL) expression is
\begin{align}
\label{eq:HTLmass}
 \left(m_{D}^{\text{HTL}}\right)^2 = 4 g^2 N_c \int \frac{d^3 p }{(2\pi)^3}\,\frac{f_{\text{BE}}(p)}{p} = \frac{N_c g^2 T^2}{3}\,,
\end{align}
with the Bose-Einstein distribution $f_{\text{BE}}(p) = (\exp(p/T)-1)^{-1}$. 

\item
In classical thermal equilibrium one has $f_{\text{BE}}(p) \rightarrow T/p$, and the discretization on an anisotropic cubic lattice with a finite lattice spacing leads to the following hard classical loop (HCL) expression
\begin{align}
\label{eq:HCLmass}
 \left(m_{D}^{\text{HCL}}\right)^2 = \frac{g^2T N_c}{2a}\left( \frac{3\Sigma}{2\pi}-1 \right),
\end{align}
where $\Sigma \approx 3.1759$ is a factor that results from the anisotropy of the lattice \cite{Moore:1999fs}.
Strictly speaking, this expression is the leading-order perturbative result in the infinite volume limit. Since we employ large lattice sizes, the latter effect can be neglected. Moreover, finite volume effects are expected to be less important than sub-leading perturbative effects that typically contribute to \eqref{eq:HTLmass} and \eqref{eq:HCLmass} at low momenta $p \lesssim m_D$. We will use \eqref{eq:HCLmass} in our figures when we plot the CYM lattice results as a function of $m_D r$.

\item
Our final possibility is to measure the Debye mass $\hat{m}_D$ directly from our thermalized CYM configurations using chromo-electric field correlators, which is explained in the previous subsection. Its temperature, and thus $\beta$ dependence is shown in \fig\ref{plot:Tmd_T0} for different lattice spacings. One generally finds that it grows with temperature. More precisely, we find an approximately linear connection to $m_{D}^{\text{HCL}}$ with 
\begin{align}
\label{eq:mdhat_md}
 \hat m_D \approx 1.42\; m_{D}^{\text{HCL}}
\end{align}
for $92 \leq \beta \leq 300$, while at lower $\beta$ deviations from a linear connection become more sizable. 

\end{enumerate}

\subsection{Parameters and scaling properties with the coupling}
\label{sec:coupling_scaling}

The coupling $g$ can be scaled out of the dynamical equations of motion \eqref{eq:classEOM} by rescaling all field amplitudes as $gA \mapsto A$ and $gE \mapsto E$. As visible in \eqref{eq:extractT}, the coupling then only enters in the combination $g^2 T$. This dimensionful scale can then be used to make all quantities dimensionless by rescaling them with appropriate powers. In the rescaled version, one then has $\langle EE \rangle_T \mapsto 1$. Note that this rescaling implies that lattice simulations only depend on the lattice size $N^3$ and on the lattice spacings $g^2Ta \equiv 2N_c/\beta$ and $g^2Ta_t$. Since we use $a_t \ll a$, temporal artefacts that may result from a dependence on $g^2Ta_t$ are suppressed, and the simulations mainly depend on 
\begin{align}
 \beta = \frac{2N_c}{g^2Ta}\,.
\end{align}
If not stated otherwise, we will write our values for $a$, $g^2 T_0$, the extracted temperature $g^2 T$ and all dimensionful variables in lattice units. We will also provide the corresponding $\beta$ values.

%%%%%%%%%%%%%%%%%%%%%%%%%%%%%%%%%%%%%%%%%%%%%%%%%%%%%%%%%%%%%
\section{Results}
\label{sec:results}
%%%%%%%%%%%%%%%%%%%%%%%%%%%%%%%%%%%%%%%%%%%%%%%%%%%%%%%%%%%%%

In this section we present the non-perturbative results from classical-statistical simulations of real-time Yang-Mills theory on cubic lattices with $N^3$ sites and the spacing length of $a$. As detailed in \se\ref{sec:theory}, the fields are initialized with initial temperature $T_0$, they thermalize by solving classical equations of motion, and the actual temperature $T$ is extracted with an error estimate from correlation functions of chromo-electric fields. 
We present the values of the imaginary part of the classical potential ${\rm Im}[V_{\rm cl}(r)]$ for $SU(3)$ extracted from the evolution of temporal Wilson loops \cite{Laine:2006ns, Laine:2007qy}. For comparison, we start this section by recalling 
the corresponding analytic results from {dimensionally regularized HTL} and the expressions from {lattice-regularized HCL} calculations of  ${\rm Im}[V_{\rm cl}(r)]$ to the second-order. We compute the HCL potential numerically for a wide range of $\beta$ values and fit the numerical HCL results to the functional form of a suitable analytically available approximation with only two parameters. The fit allows us to extract the $\beta$ dependence of the parameters of the HCL potential, including an estimate for the large-$\beta$ limit with $\beta \to \infty$. 
We then discuss our numerical results from classical Yang-Mills simulations for a wide range of $\beta$ values, and compare them with previously published data of ${\rm Im}[V_{\rm cl}(r)]$ obtained from lattice calculations, with simulations in $SU(2)$ theory and with HCL results. Our main simulation results are also summarized in tables in \app\ref{app:tables}.

\subsection{Hard Thermal Loop result}
\label{sec:HTL_V}

In Ref.~\cite{Laine:2006ns} the authors derived an expression for the imaginary part of the heavy-quark potential to leading-order in the strong coupling constant using the continuum hard thermal loop framework and dimensional regularization.  Their final result could be expressed compactly as
\be
{\rm Im}[V^{(2)}(r)] = - \frac{C_F g^2 T}{4\pi} \phi\left(m_{D}^{\text{HTL}}\, r\right) \, .
\label{htlimv}
\ee
where $C_F = (N_c^2 -1)/2N_c\,$, 
\be 
\phi\left( x \right) \equiv 2 \int_0^{\infty} {\rm d}z \frac{z}{\left(1+z^2 \right)^2}\left[ 1- \frac{\sin\left(zx\right)}{zx}\right] ,
\label{eq:phidef}
\ee
and $m_{D}^{\text{HTL}}$ given by the continuum hard thermal loop Deybe mass \eqref{eq:HTLmass}.  Note that in the large-$r$ limit one has $\lim_{r \rightarrow \infty} \phi(r) = 1$.  As a result the asymptotic value of the imaginary part in this case is $-C_F g^2 T/4\pi$.

%%%%%%====FIGURE======%%%%%%%%%%%%%%%%%%%%%%%%%%%%%
\begin{figure}[t]
\centerline{
\includegraphics[width=0.48\linewidth]{\pToFigs/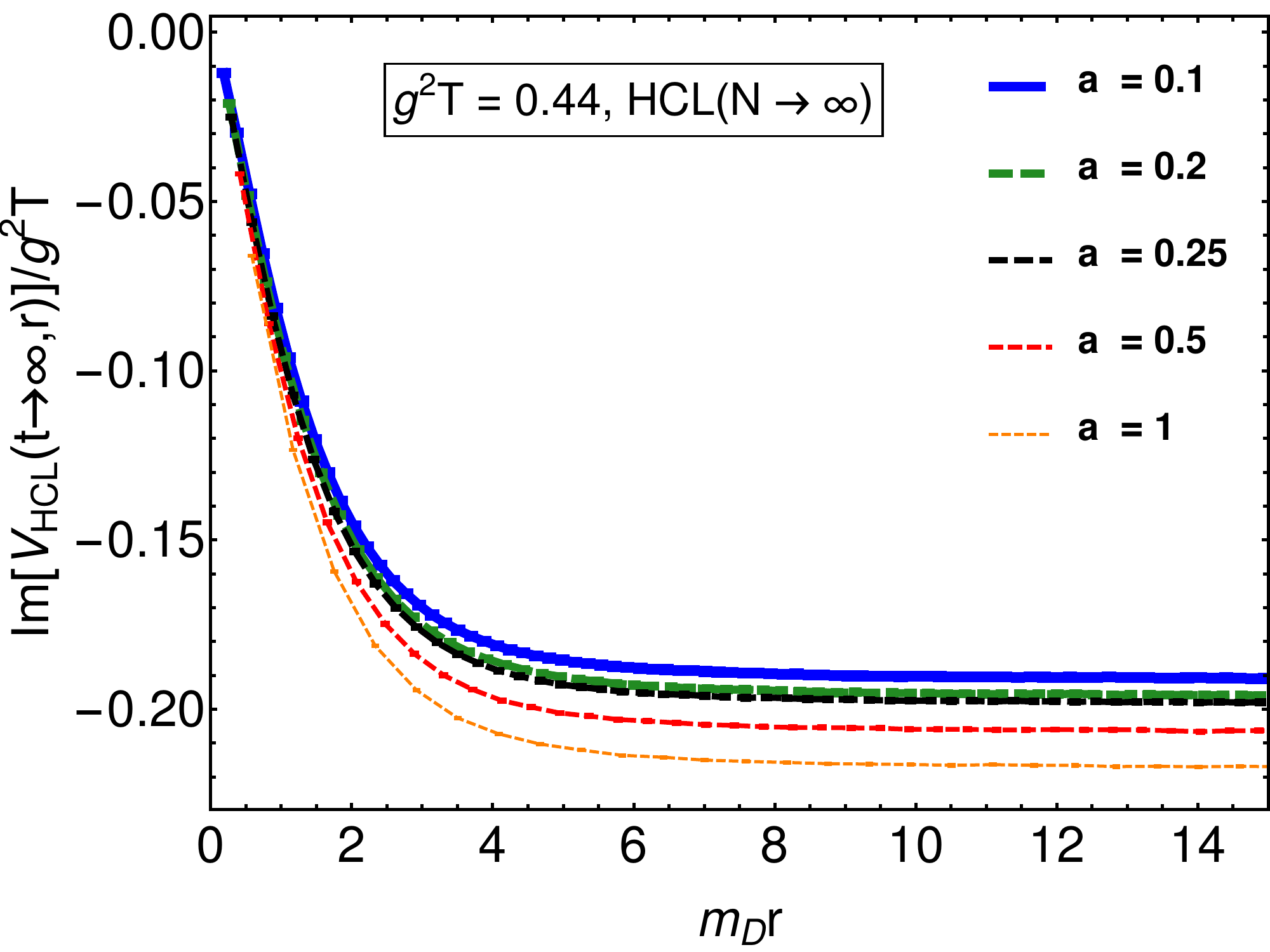}
$\;\;\;$
\includegraphics[width=0.48\linewidth]{\pToFigs/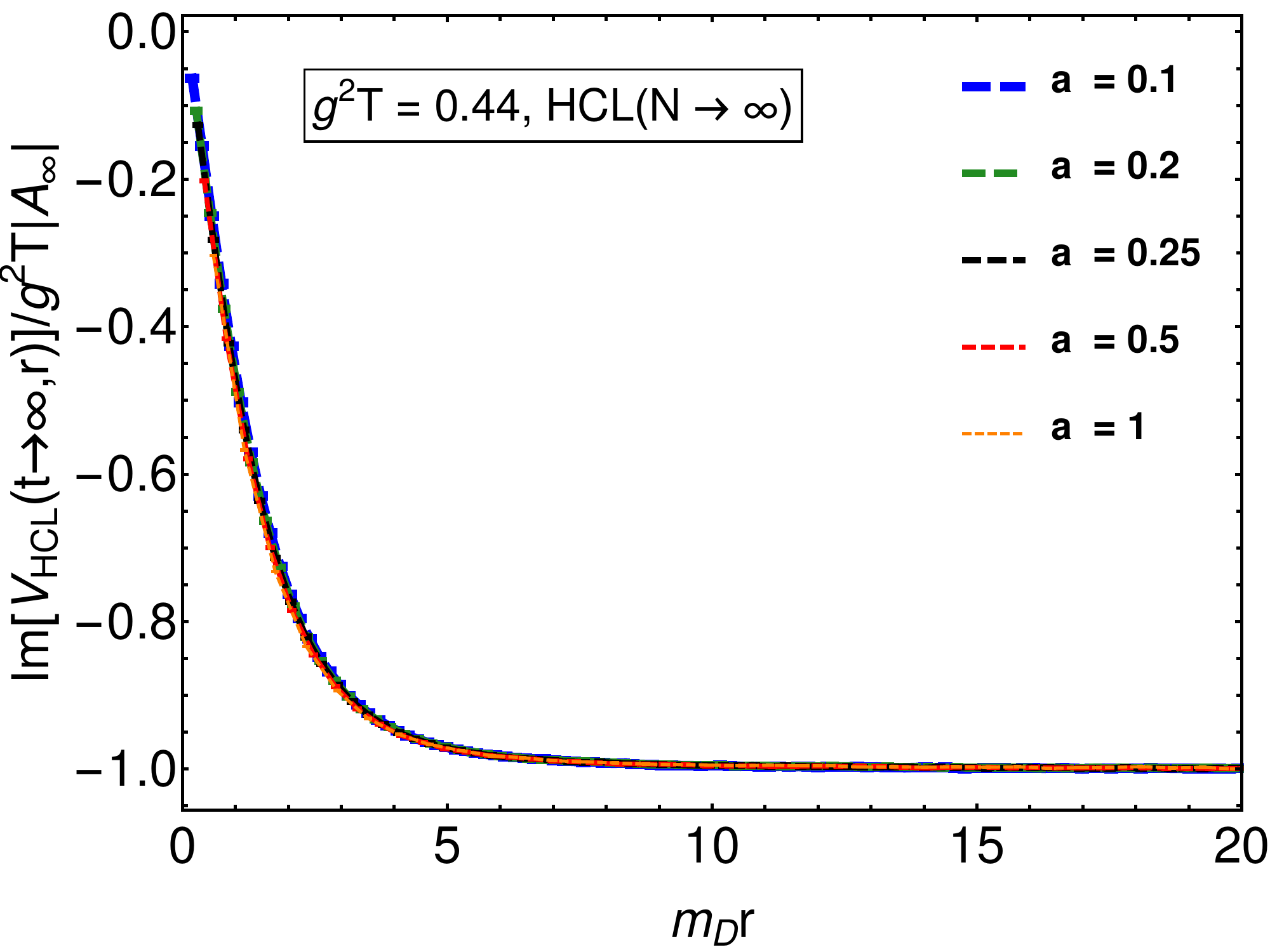}
}
\caption{(Left) Large-time limit of ${\rm Im}[V^{(2)}_{\rm cl}(r)]/g^2T$ for $g^2T = 0.44$ and $a \in \{1, 0.5, 0.25, 0.2, 0.1 \}$ corresponding to $\beta \in \{13.6, 27.3, 54.6, 68.2, 136.4\}$, calculated using the infinite-volume hard classical loop (HCL) approximation in \eqref{hclimv}. These numerical results are represented by points while fits using the function \eqref{Aphi} are shown as lines.
(Right)  Same as in the left panel, but divided by the asymptotic value $A_\infty \equiv \lim_{r \rightarrow \infty} {\rm Im}[V^{(2)}_{\rm cl}(r)]/g^2T$.}
\label{plot:hcl_1}
\end{figure}
%%%%%%%%%%%%%%%%%%%%%%%%%%%%%%%%%%%%%%%%%%%%%%%%%%%%

\subsection{Hard Classical Loop results}
\label{sec:HCL_results}

The result for the imaginary part of the classical potential in the infinite volume and infinite time limit ($N \rightarrow \infty,\ t \rightarrow \infty$) from second-order perturbation theory regularized on a cubic lattice (HCL: Hard Classical Loop) of size $\left(aN\right)^3$  is calculated in Ref.~\cite{Laine:2007qy} and is given by
\be 
{\rm Im}[V^{(2)}_{\rm cl}(r)]/g^2T = -\frac{\pi C_F N_c^2}{\beta} \int_0^1 {\rm d}^3{\mathbf{x}}\frac{1-\cos\left( \pi x_3r/a \right)}{ (\tilde{\mathbf{x}}^2 + N_c^2\Sigma/\pi \beta )^2} \int_{-1}^1 {\rm d}^3{\mathbf{y}} \frac{\delta\left( {\tilde{\mathbf{x}} \cdot \mathring{\mathbf{y}} } \right)}{ \left(\tilde{\mathbf{y}}^2\right)^{1/2} }, \label{hclimv}
\ee
where the accented variables are defined as 
\be 
\tilde{x}_i \equiv 2\sin\left( \frac{\pi x_i}{2}\right), \qquad \mathring{x}_i \equiv \sin\left(\pi x_i\right), \qquad x_i \in \left( -1, 1 \right).
\ee
%$C_A = N_c$, 
%$C_F = (N_c^2 -1)/2N_c$ and $\beta = 2N_c/g^2aT$. 
We perform the numerical integration necessary in \eqref{hclimv} using the VEGAS algorithm \cite{vegas}, as implemented in the CUBA library \cite{cuba}. 
The results for $g^2T = 0.44$ and $a \in \{1, 0.5, 0.25, 0.2, 0.1 \}$ 
corresponding to $\beta \in \{13.6, 27.3, 54.6, 68.2, 136.4\}$ are shown as points in the left panel of Fig.~\ref{plot:hcl_1} as functions of $m_D r$ with $m_D$ given by the HCL Debye mass \eqref{eq:HCLmass}. 
As can be seen from this panel, the asymptotic large-$r$ value of ${\rm Im}[V_{\rm HCL}]$ depends on $\beta$, however, the small-$r$ behavior is similar for all values of $\beta$ shown. We find that the numerical HCL curves are fit very well by the functional form
\be 
{\rm Im}[V^{(2)}_{\rm cl}(r)] = g^2T A_{\infty}\phi\left(B\,m_D r\right) .
\label{Aphi}
\ee
with $\phi(x)$ given in Eq.~\eqref{eq:phidef}. 
This functional form is motivated by the analytic leading-order perturbative HTL result \eqref{htlimv} for which $A^\text{HTL}_{\infty} = -C_F/4\pi$ and $B^\text{HTL}=1$ with the HTL calculated $m_D$. The fitted curves are shown as lines in \fig\ref{plot:hcl_1} and nicely coincide with the numerically integrated results. 

%%%%%%====FIGURE======%%%%%%%%%%%%%%%%%%%%%%%%%%%%%
\begin{figure}[tp]
\centerline{
\includegraphics[width=0.48\linewidth]{\pToFigs/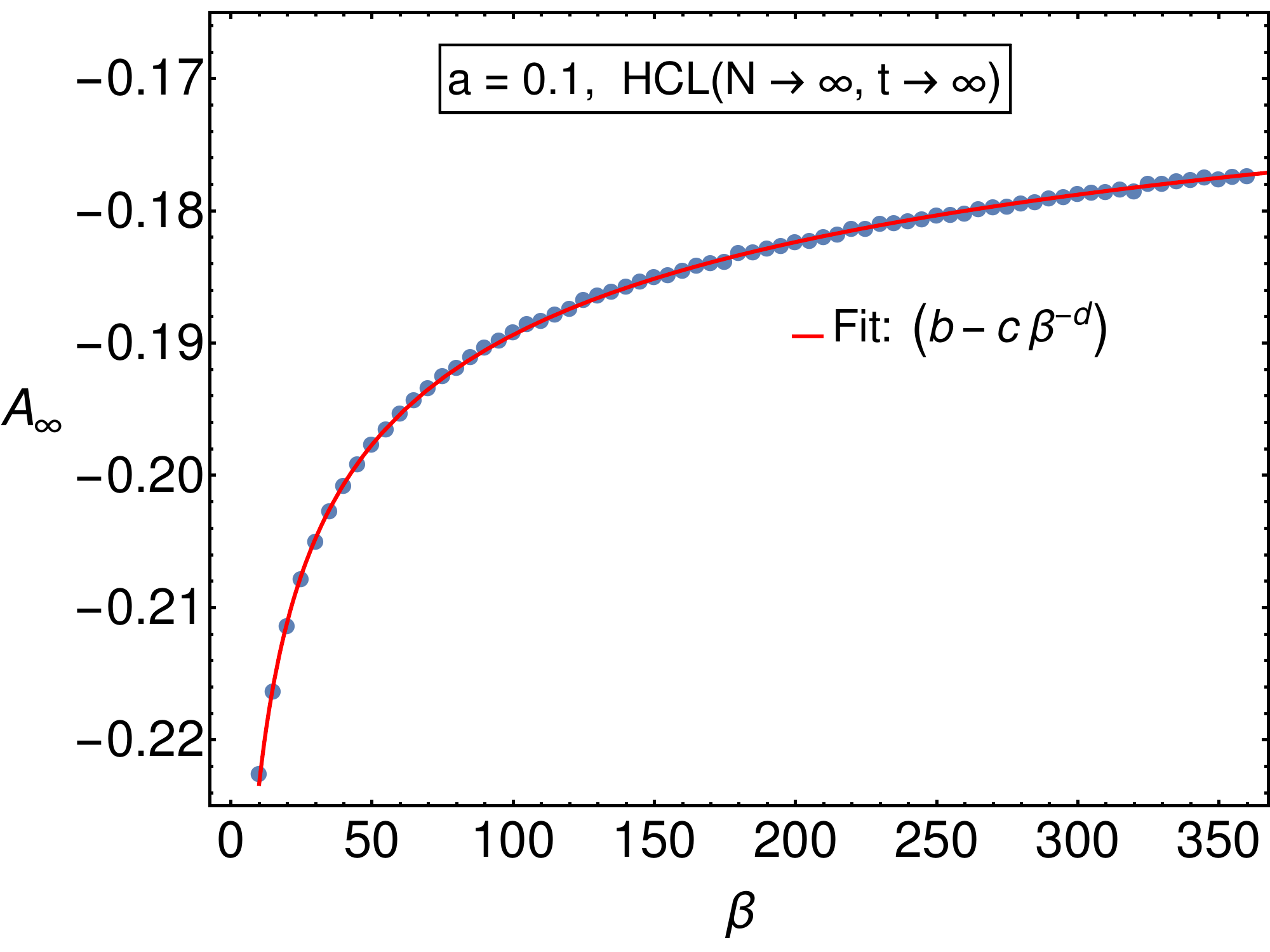}
$\;\;\;$
\includegraphics[width=0.48\linewidth]{\pToFigs/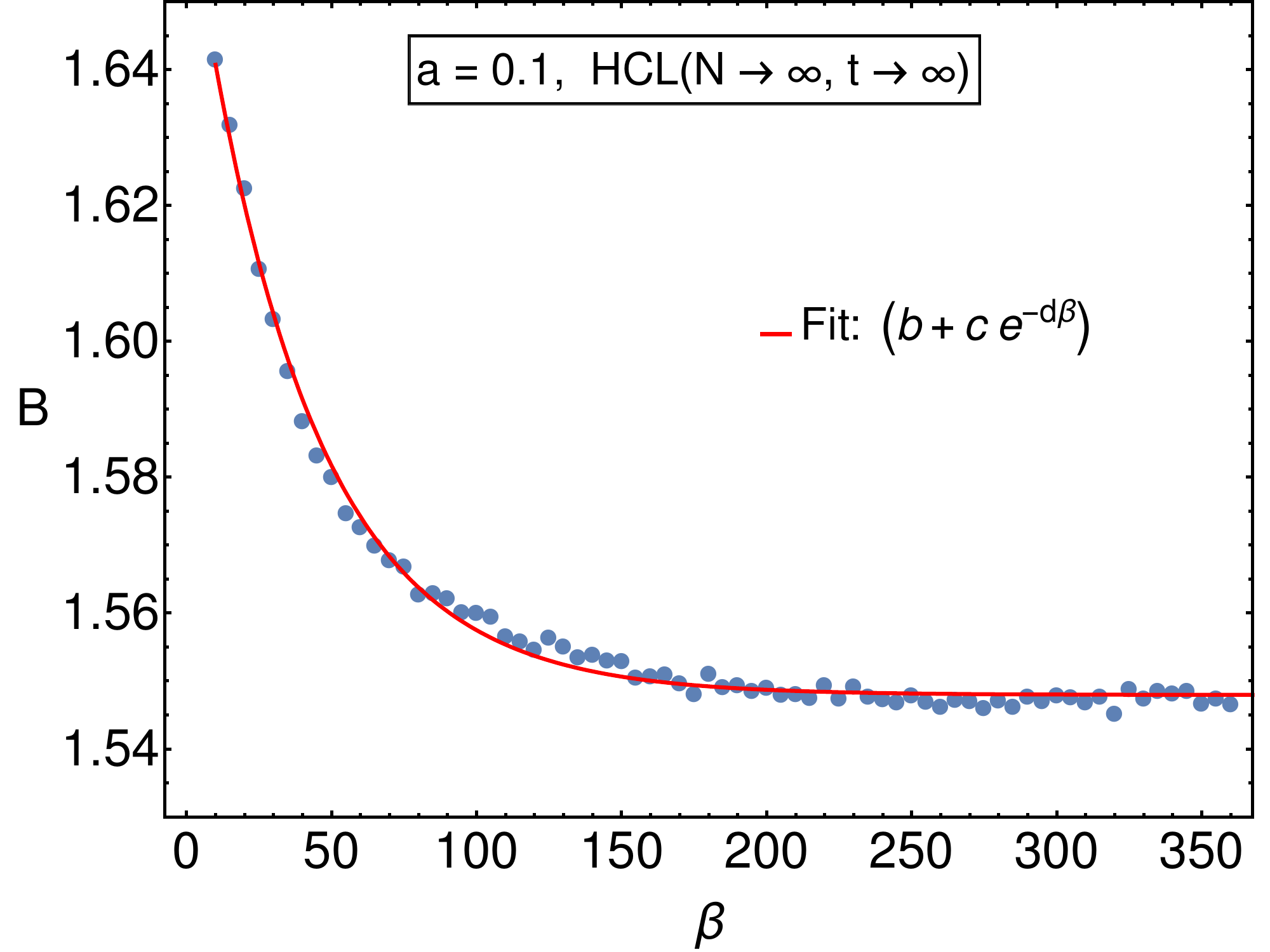}
}
\caption{Fit values $A_\infty^{\text{HCL}}$ (left) and $B^{\text{HCL}}$ (right) of the fit function \eqref{Aphi} to the classical potential ${\rm Im}[V^{(2)}_{\rm cl}(r)]/g^2T$ computed within HCL for a wide parameter range $\beta = 10 \rightarrow 360$ for $a = 0.1$. Fits to the dependence of these parameters on $\beta$ are shown as red curves, with the chosen form of the fit functions indicated in the label. The asymptotic values are $A_\infty^{\text{HCL}}(\beta {\to} \infty)=-0.144$ and $B^{\text{HCL}}(\beta {\to} \infty)=1.55$.}
\label{plot:hcl_AB}
\end{figure}
%%%%%%%%%%%%%%%%%%%%%%%%%%%%%%%%%%%%%%%%%%%%%%%%%%%%

To make the agreement better visible, we show in the right panel of Fig.~\ref{plot:hcl_1} the same results divided by their corresponding asymptotic values $A_\infty^{\text{HCL}}$. All curves fall on top of each other, demonstrating that ${\rm Im}[V^{(2)}_{\rm cl}(r)]/(g^2TA_\infty^{\text{HCL}})$ is well described by $\phi\left(B^{\text{HCL}}\,m_{D}^{\text{HCL}}\, r\right)$. This also shows that, while fitted values of $A_{\infty}^{\text{HCL}}$ depend on $\beta$, this functional form and, in particular, the parameter $B^{\text{HCL}}$ is barely sensitive to $\beta$ in the considered range of $\beta$ values.

We show the fitted parameters $A_\infty^{\text{HCL}}$ and $B^{\text{HCL}}$ in \fig\ref{plot:hcl_AB} for a wide range $10 \leq \beta \leq 360$. Consistent with our earlier observations, $B^{\text{HCL}}$ becomes practically constant at large $\beta$, while $A_\infty^{\text{HCL}}$ is still growing for the values considered here. To better understand the functional form of the parameters and to extrapolate to larger $\beta$, we have performed fits with different functional forms. The functions that seem to describe the results in the best way are 
\begin{align}
 A_\infty^{\text{HCL}}(\beta) &\approx -0.144 - 0.139\, \beta^{-0.24} \nonumber \\
 B^{\text{HCL}}(\beta) &\approx 1.548 + 0.12\,e^{-0.025 \beta}\,,
\end{align}
and are included into \fig\ref{plot:hcl_AB} as red continuous lines.%
\footnote{We tried both functional forms for each of the parameters $A_\infty$ and $B$ and found that these forms shown in the plots are preferred by far.}
The fit functions allow us to estimate the asymptotic values
\begin{align}
\label{eq:AB_HCL_contlimit}
 A_\infty^{\text{HCL}}(\beta {\to} \infty) \approx -0.144 = -0.108\,C_F\,, \qquad B^{\text{HCL}}(\beta {\to} \infty) \approx 1.548 \,. 
\end{align}
Since the large-$\beta$ limit corresponds to the limit $a \to 0$, its existence is very nontrivial for an observable in classical thermal lattice field theory due to the Rayleigh-Jeans divergence. Our analysis suggests that such a limit exists for ${\rm Im}[V^{(2)}_{\rm cl}(r)]$. This was possible in our case because, different from previous studies, we studied the HCL potential as a function of $m_{D}^{\text{HCL}}\, r$ rather than any other combination of the lattice spacing or the temperature. Since $m_{D}^{\text{HCL}}$ diverges for small $a$, our parametrisation hides the divergence and leaves the functional form and the remaining parameters finite. It is interesting to compare the extrapolated parameter values \eqref{eq:AB_HCL_contlimit} to the corresponding HTL values in dimensional regularization of \se\ref{sec:HTL_V}. As mentioned before, these are
\begin{align}
\label{eq:AB_HTL}
 A^\text{HTL}_{\infty} \approx -0.08\, C_F\,, \qquad B^\text{HTL}=1\,.
\end{align}
The $A_\infty$ values are quite close, with the estimated HCL value $|A_\infty^{\text{HCL}}(\beta {\to} \infty)|$ being $35\%$ larger than the corresponding HTL value. Note here that our extrapolation is expected to have a large uncertainty for $A_\infty^{\text{HCL}}$ for $\beta \to \infty$ since its value is still quite rapidly changing in the considered $\beta$ interval, as seen in \fig\ref{plot:hcl_AB}. While the fit works surprisingly well over a large $\beta$ interval, which gave us confidence to write a limiting value, our analysis is not a proof that $|A_\infty^{\text{HCL}}(\beta {\to} \infty)|$ is indeed finite but rather an observation from this fit function. This is different for the parameter $B$, that agrees well with its estimated large-$\beta$ limit for $\beta \gtrsim 200$. Since $B$ multiplies the divergent Debye mass, the product $B\, m_{D}^{\text{HCL}}$ could be interpreted as a \changeflag{replacement for} the lattice regularized mass.

%%%%%%%%%%%%%%%%%%%%%%%%%%%%%%%%%%%%%%%%%%%%%
%% Results of the fits:
%%%%%%%%%%%%%%%%%%%%%%%%%%%%%%%%%%%%%%%%%%%%%
%
% For $A_\infty$
% {b -> -0.14429, c -> 0.138575, d -> 0.243843}
% Errors: {0.000786122, 0.000306184, 0.00398507}
%% CHECKED: Derivative of $A_\infty$ points approaches zero
%
% For $B$
% {b -> 1.54794, c -> 0.119445, d -> 0.0252848}
% Errors: {0.000245528, 0.00162114, 0.000471163}
%%%%%%%%%%%%%%%%%%%%%%%%%%%%%%%%%%%%%%%%%%%%%

\subsection{Nonperturbative CYM simulation results for the classical Wilson loop}

\begin{figure}[t]
\centerline{
\includegraphics[width=0.48\linewidth]{\pToFigs/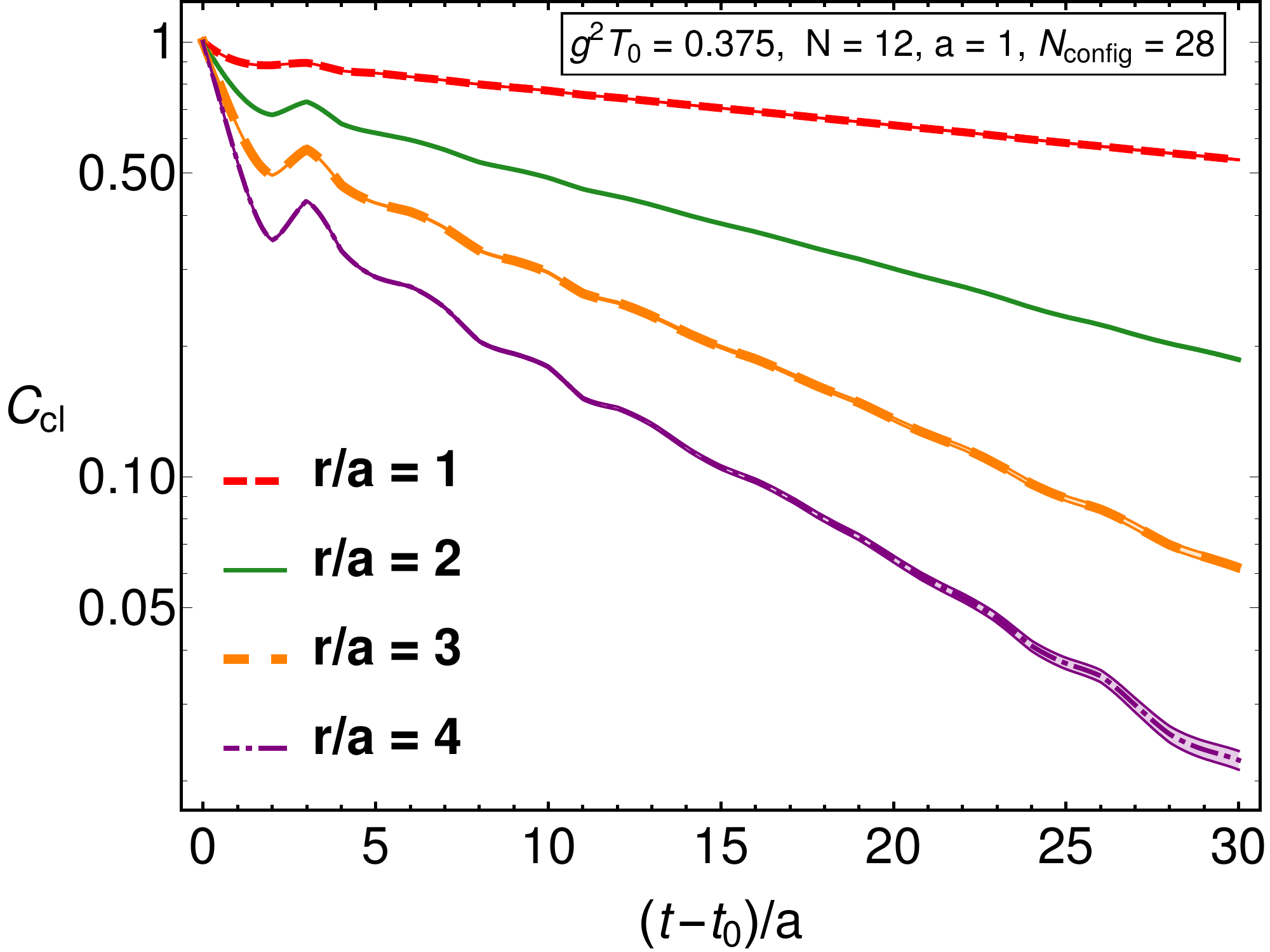}
$\;\;\;$
\includegraphics[width=0.48\linewidth]{\pToFigs/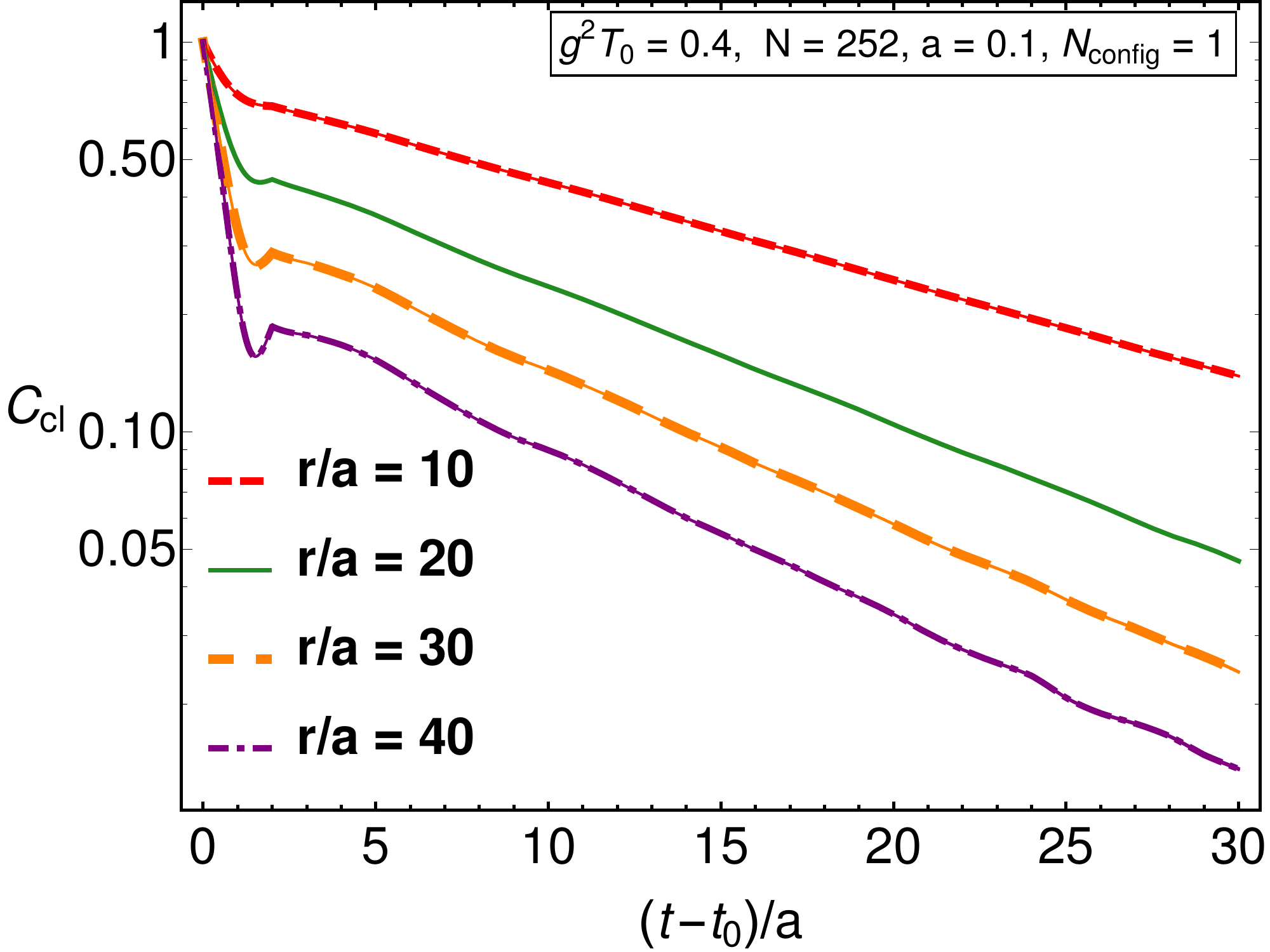}
}
\caption{Examples of the classical Wilson loop as a function of time for lattice spacing $a = 1$ and spatial separation values $r/a = \{1,\ 2,\ 3,\ 4\}$ (left), and for lattice spacing $a = 0.1$ and spatial separation values $r/a = \{10,\ 20,\ 30,\ 40\}$ (right).
} 
\label{plot::Ccl_example}
\end{figure}

We now turn to the results of our CYM lattice simulations.  To extract the imaginary part of the classical potential ${\rm Im}[V_{\rm cl}(r)]$ from non-perturbative lattice simulations, we follow the procedure introduced in Ref.~\cite{Laine:2007qy} and revisited in \se\ref{sec:theory_potential}. We consider rectangular Wilson loops of different spatial size $r = n a$ and temporal length $t = n_t a_t$.%
\footnote{Since we work in temporal axial gauge with $U_0 = \mathbb{1}$, the temporal edges of the Wilson loops are trivial and the loops reduce to line-line correlators.} 
We use a temporal lattice spacing of $a_t = a/100$ for all results reported in this paper. The imaginary part of the classical potential ${\rm Im}[V_{\rm cl}(r)]$ at separation $r$ for a heavy quarkonium system is calculated from the large-time slope of the logarithm of the Wilson loop $C_{\rm cl}(r,t)$, averaged over lattice sites, orientations and configurations. The time evolution of $C_{\rm cl}(r,t)$ is shown for two example cases in Fig.~\ref{plot::Ccl_example}.

%%%%%%====FIGURE======%%%%%%%%%%%%%%%%%%%%%%%%%%%%%
%%%%%%%%%%%%%%%%%%%%%%%%%%%%%%%%%%%%%%%%%%%%%%%%%%%%

In the left panel of Fig.~\ref{plot::Ccl_example} we present typical results obtained on a small lattice which is averaged over $N_\text{config}=28$ classical-statistical configurations.  The bands indicate the uncertainty in the mean obtained by dividing the standard deviation of the mean value by $\sqrt{N_\text{config} N_{\text{iso}} N_\text{start.points}}$.  As indicated from this last expression, we exploit the isotropy and homogeneity of the lattice in order to obtain increased statistics, where we average over two different Wilson loop orientations 
and use $N_\text{start.points} = N^3$ possible spatial starting points from which we can measure the Wilson loop.
As can be seen from this panel, one cannot perform the fits to extract the logarithmic slope at early times.  In practice, one must wait a sufficient amount of time beyond the thermalization time ($t_0$).  Typically, we use a fit window that 
encompasses $(t-t_0)/a > 10$.  We find that as long as one uses starting times for the fit window which satisfy this constraint, there is only a very small variation in the extracted logarithmic slopes.

In the right panel of Fig.~\ref{plot::Ccl_example} we present typical results obtained on a large lattice in which we only take into account one configuration for the initial fields.  As this panel demonstrates, when using large lattices, the uncertainty is decreased due to the much larger number of starting points that can be used to measure the Wilson loop.  As a result, on large lattices one does not need to sample as many initial configurations, which helps to reduce the run time required to extract statistically accurate results.  We have explicitly checked that the uncertainty bands obtained with one initial configuration are consistent with results obtained using larger number of initial configurations but a smaller lattice size.

%%%%%%====FIGURE======%%%%%%%%%%%%%%%%%%%%%%%%%%%%%
\begin{figure}[t]
\centerline{
\includegraphics[width=0.7\linewidth]{\pToFigs/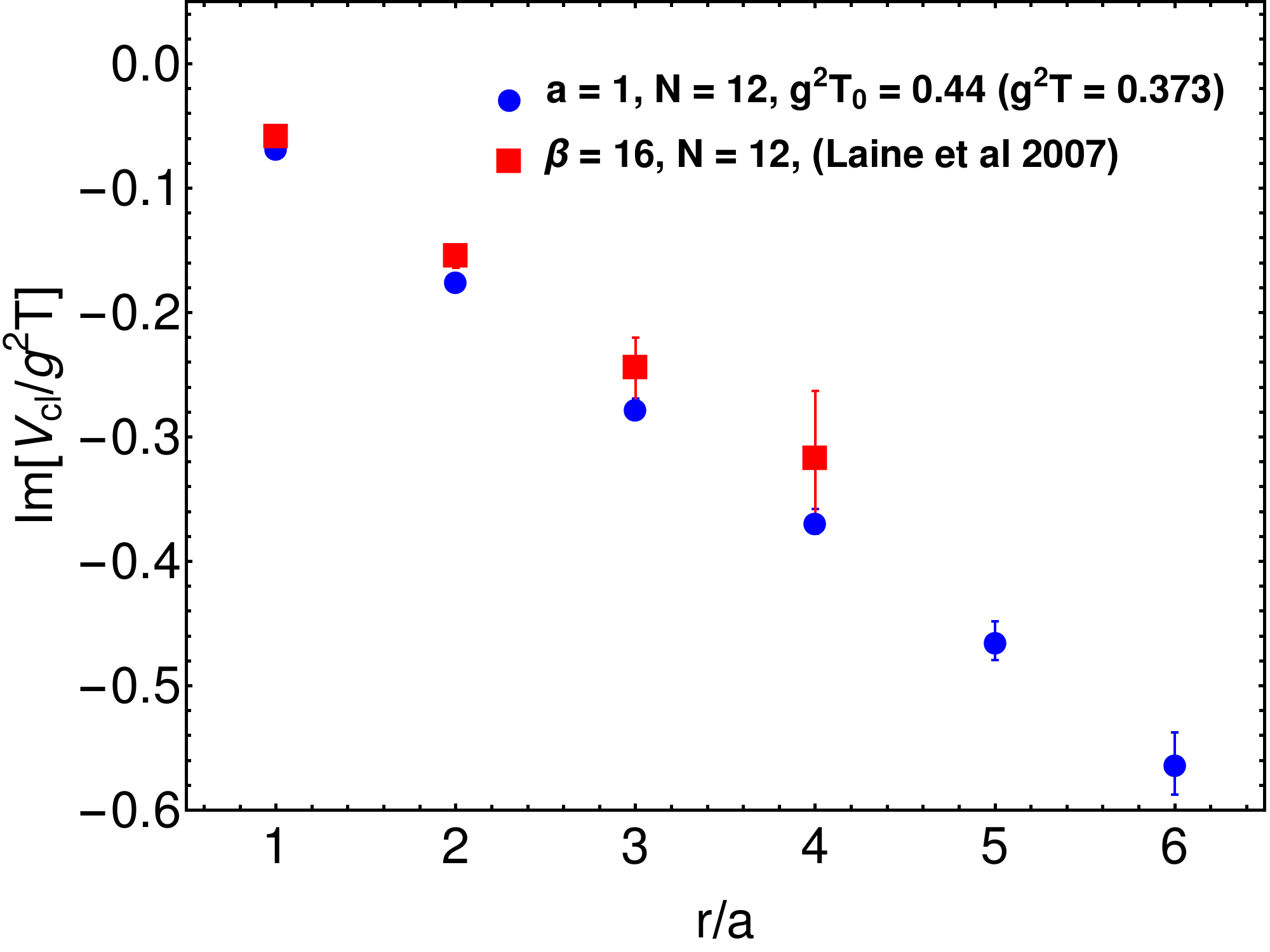}
}
\caption{${\rm Im}[V_{\rm cl}(r)]/g^2T$ as a function of $r/a$ for $\beta = 16$.  The blue filled disks are the results obtain herein averaged over 28 configurations.  The red filled squares are the results reported by Laine et al.\ in Ref.~\cite{Laine:2007qy}.} 
\label{plot:beta16}
\end{figure}
%%%%%%%%%%%%%%%%%%%%%%%%%%%%%%%%%%%%%%%%%%%%%%%%%%%%

\subsubsection{Benchmark results for $\beta = $ 16 and $a=1$}

In Fig.~\ref{plot:beta16} we compare our results to the previously available ones from Ref.~\cite{Laine:2007qy} for $\beta = 16$. We set the parameter $g^2 T_0 = 0.44$ for the initial distribution so that the resulting temperature extracted from $\langle EE \rangle$ correlators is $g^2 T \approx 0.373$, which is close to the value $g^2 T = 0.375$ for the corresponding $\beta = 16$ when setting $a = 1$, $g^2 = 1$.
The error bars indicated in Fig.~\ref{plot:beta16} include uncertainties from averaging the Wilson loops over the starting points, orientations, classical configurations, uncertainties coming from the fitting residuals, and uncertainties from the temperature extraction. As can be seen from this Figure, our results are in agreement within uncertainties with the reported results from Ref.~\cite{Laine:2007qy}.

%%%%%%====FIGURE======%%%%%%%%%%%%%%%%%%%%%%%%%%%%%
\begin{figure}[t]
\centerline{
\includegraphics[width=0.48\linewidth]{\pToFigs/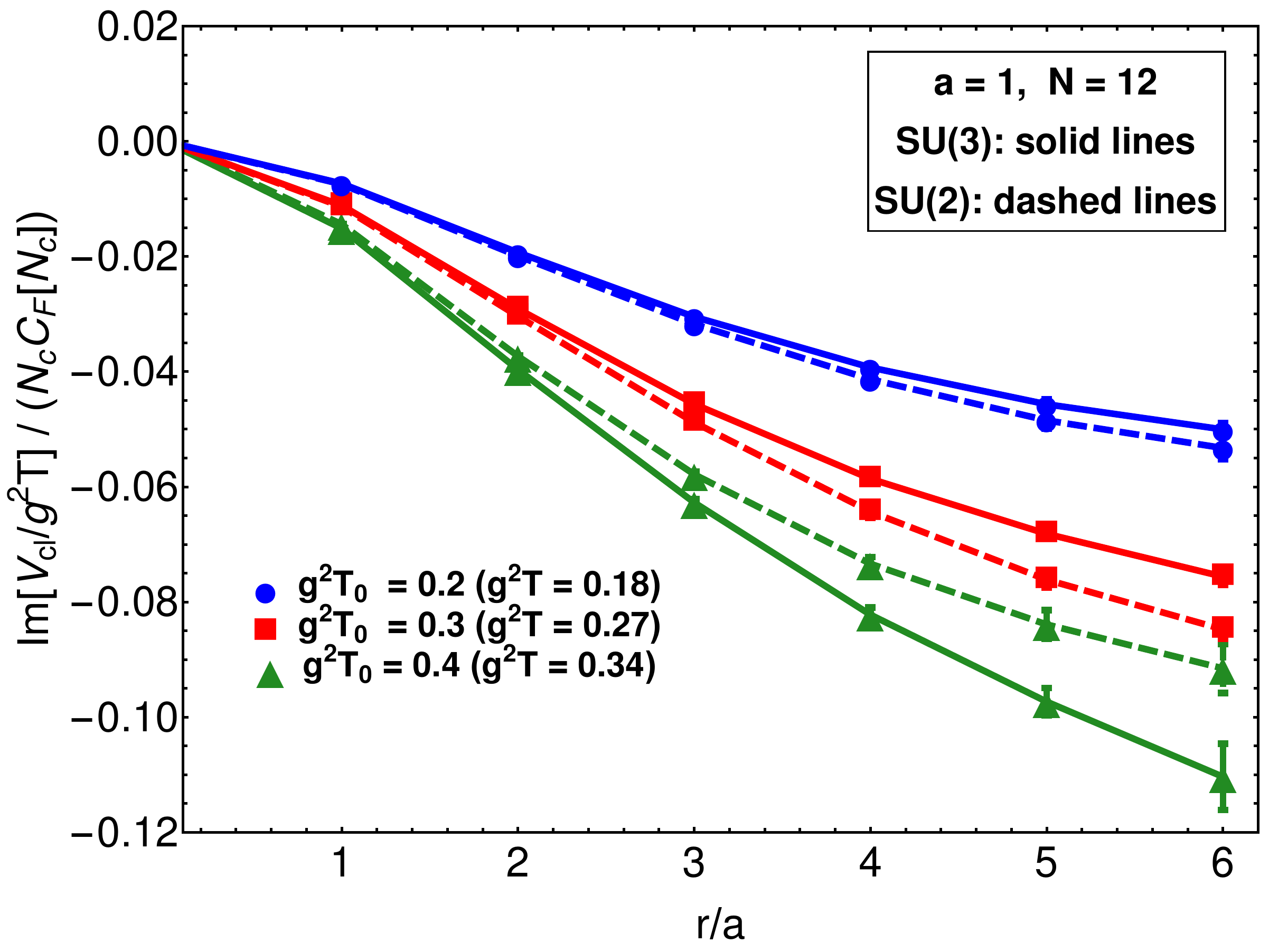}
$\;\;\;$
\includegraphics[width=0.48\linewidth]{\pToFigs/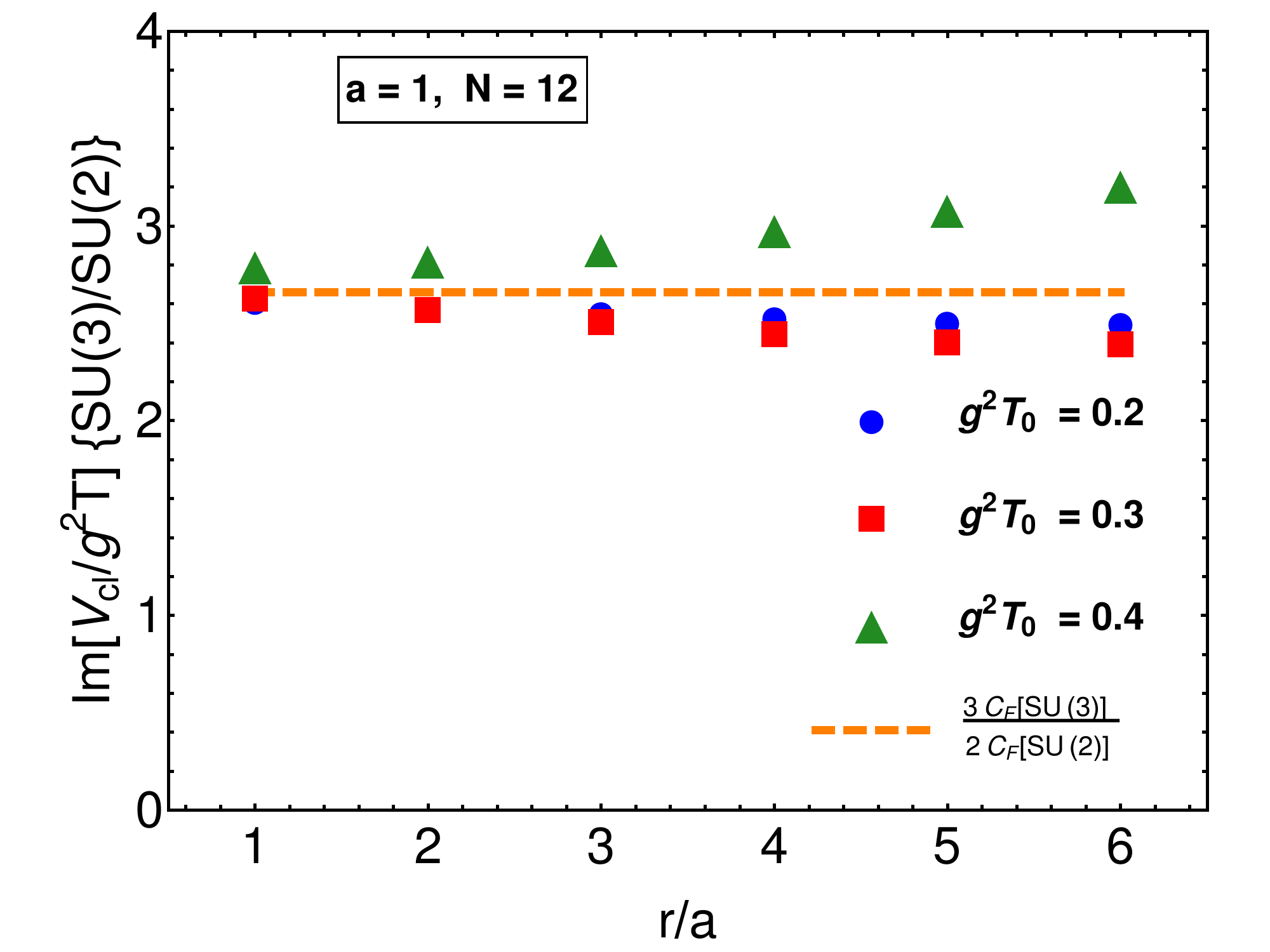}
}
\caption{{\em Left:} ${\rm Im}[V_{\rm cl}(r)]/g^2T$ extracted from $SU(2)$ and $SU(3)$ lattice gauge theories simulated with the same values $T_0$. The given values of $T$ correspond to $SU(3)$. The potential is divided by $C_F N_c$ of the respective gauge group. {\em Right:} The ratios of $SU(3)$ to $SU(2)$ results for ${\rm Im}[V_{\rm cl}(r)]/g^2T$ for the same simulations. The orange dashed line corresponds to $8/3$.}
\label{plot:su2_1_2}
\end{figure}
%%%%%%%%%%%%%%%%%%%%%%%%%%%%%%%%%%%%%%%%%%%%%%%%%%%%

\subsubsection{Comparing results for SU(3) vs.\ SU(2)}
\label{sec:SU3_vs_SU2}

To check for the $N_c$ dependence of the $SU(N_c)$ gauge theories, we perform here similar lattice calculations for $SU(3)$ and $SU(2)$ systems and extract ${\rm Im}[V_{\rm cl}(r)]/g^2T$ for different values of $T_0$. In \fig\ref{plot:su2_1_2} we use the same lattice parameters $a=1$ and $N=12$ as we did in \fig\ref{plot:beta16} and compare the simulations for the values $g^2 T_0 = \{0.2,0.3,0.4\}$. In the left panel of \fig\ref{plot:su2_1_2}, the values of the ratio ${\rm Im}[V_{\rm cl}(r)]/(g^2T\,C_F N_c)$ are shown for the two gauge groups as functions of $r/a$. One finds that they agree well at low $r/a$ while the deviations grow with the distance. However, the agreement improves for smaller temperatures (larger $\beta$ values). 

In the right panel, the ratio of the imaginary parts of the potentials of the $SU(3)$ and $SU(2)$ theories, $({\rm Im}[V_{\rm cl}(r)]/(g^2T)[SU(3)]) / ({\rm Im}[V_{\rm cl}(r)]/(g^2T)[SU(2)])$, is plotted as a function of $r/a$. The horizontal dashed line corresponds to $(N_c C_F[SU(3)]) / (N_c C_F[SU(2)])=8/3$, which is the ratio of their $N_c C_F$ values. One observes that the data points are close to this value for all $r$ and $T_0$, which shows
that the imaginary part of the potential in our lattice simulations admits Casimir scaling
\begin{align}
\label{eq:V_propto_CN}
 {\rm Im}[V_{\rm cl}(r)] \propto C_F N_c\,.
\end{align}
This is similar to the perturbative HCL expression \eqref{hclimv} where due to $\beta \propto N_c$, one also finds this relation \eqref{eq:V_propto_CN}. 

The deviations visible in the right panel of \fig\ref{plot:su2_1_2} decrease with increasing $\beta$ (smaller temperature) and with decreasing distance, as we have observed in the left panel. 
An important source for these deviations is that we compare here simulations with the same $g^2 T_0$ parameter and not with the same temperature $g^2 T$. As visible in the left panel of \fig\ref{plot:Tmd_T0} for $a=1$ and $SU(3)$ theory, the extracted temperatures deviate stronger at larger $g^2 T_0$. The deviations in $SU(2)$ theory are qualitatively similar but can lead to quantitatively different values of $g^2 T$, which implies that we compare the $SU(N_c)$ theories effectively at slightly different temperatures. This effect is reduced when going to smaller temperatures or smaller lattice spacings (larger $\beta$ values). 
Moreover, as we will show below, the short-distance behavior is less sensitive to the temperature, which is the reason why the deviations decrease at lower $r/a$.

\subsubsection{SU(3) simulation results for different $\beta$}

\begin{figure}[t]
\centerline{
\includegraphics[width=0.8\linewidth]{\pToFigs/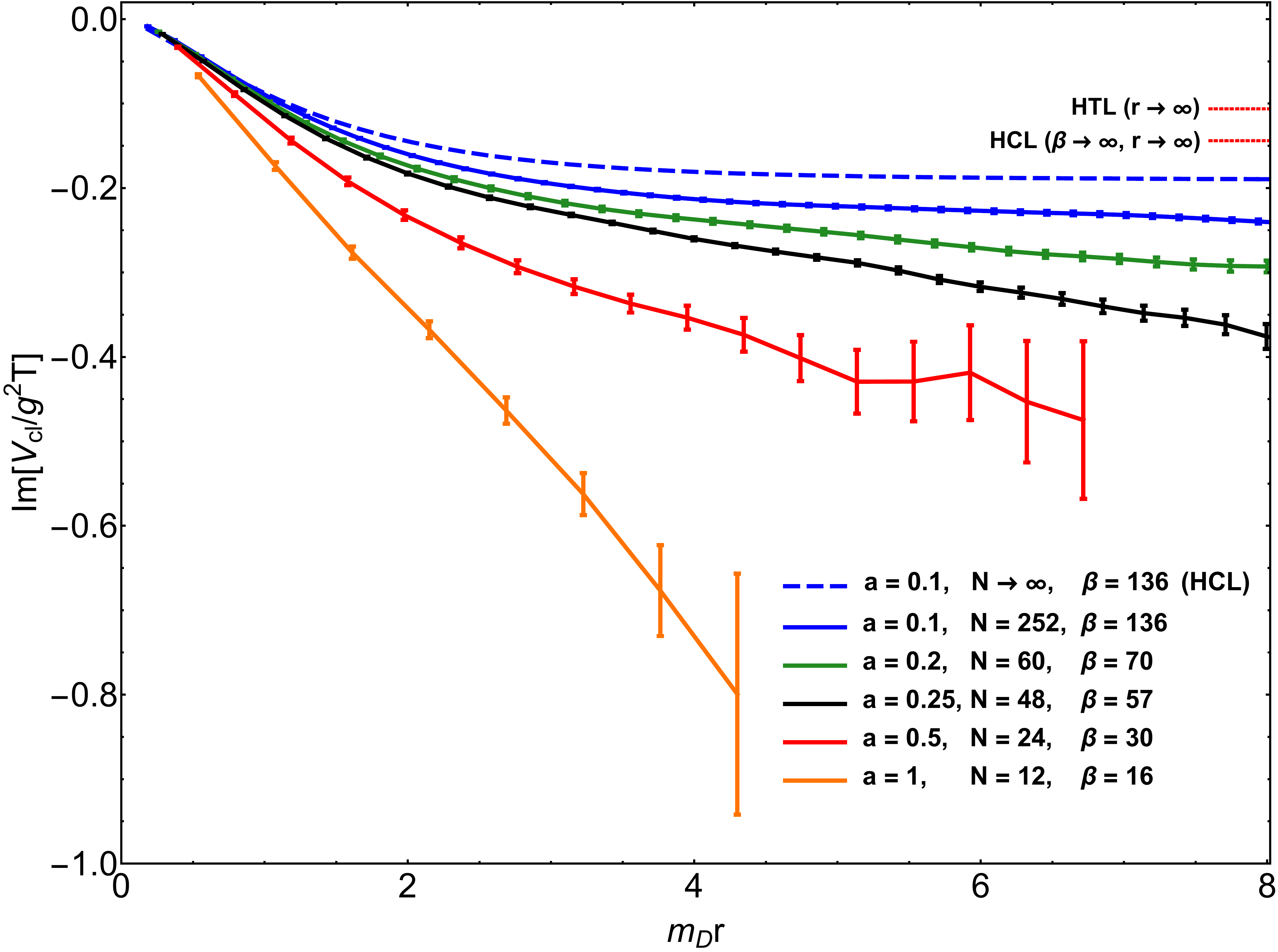}
}
\caption{ ${\rm Im}[V_{\rm cl}(r)]/g^2T$ for different lattice spacing from nonperturbative Wilson loops (solid lines) and from HCL perturbation theory for $a=0.1$ (blue dashed line). The temperature values (extracted from $EE$ correlations) are slightly different for different values of $a$, and we write the corresponding $\beta$ values instead. The $r \rightarrow \infty$ limits of HCL for the continuum extrapolation $\beta \to \infty$ \eqref{eq:AB_HCL_contlimit} and of the dimensionally-regularized HTL perturbation theories \eqref{eq:AB_HTL} are indicated by short dotted red lines on the right axis for comparison.  The distance in the horizontal axis is scaled using the HCL Debye mass \eqref{eq:HCLmass} for $m_D(a)$.} 
\label{plot:imv_mDr}
\end{figure}

We performed CYM lattice simulations with $g^2 T_0 = 0.44$ and $a = \{1,\ 0.5,\ 0.25,\ 0.2\}$ keeping $aN = 12$ fixed and also for $g^2 T_0 = 0.45$ and $a =0.1$ with $N = 252$. Our results for ${\rm Im}[V_{\rm cl}(r)]/g^2T$ are shown in Fig.~\ref{plot:imv_mDr} as points connected by solid colored lines, and the respective $\beta$ values are written in the figure. 
The $a = 0.1$ CYM lattice simulation results are compared with the corresponding results (same $a$ and $T$) obtained from HCL perturbation theory \eqref{hclimv}, included as a blue dashed curve. For both our CYM lattice results and the HCL results, 
${\rm Im}[V_{\rm cl}(r)]/g^2T$ is plotted as a function of $m_D r$,
where we use the leading-order Debye mass $m_{D}^{\text{HCL}}$ calculated in HCL theory \eqref{eq:HCLmass}. 
On the right edge of the figure we indicate the asymptotic $r \to \infty$ values obtained using both the dimensionally-regularized continuum HTL result \eqref{htlimv} and the extrapolated $N,\beta \rightarrow \infty$ HCL result with the parameters \eqref{eq:AB_HCL_contlimit}.

%%%%%%====FIGURE======%%%%%%%%%%%%%%%%%%%%%%%%%%%%%
%%%%%%%%%%%%%%%%%%%%%%%%%%%%%%%%%%%%%%%%%%%%%%%%%%%%

As can be seen from this Figure (\fig\ref{plot:imv_mDr}), the CYM lattice results seem to \prevchangeflag{approach a finite large-$\beta$ form}
when plotted as a function of $m_D r$, with the CYM results with $\beta=136$ even overlapping with the corresponding HCL curve at $m_D r \lesssim 1$.
At large values of $r$ our CYM simulation results approach the corresponding HCL curve for the respective $\beta$ value. For the largest $\beta = 136$ in the figure, we know from the left panel of \fig\ref{plot:hcl_AB} that the amplitude of the potential at large $r$ is still quite far away from the extrapolated value. However, \fig\ref{plot:imv_mDr} indicates that \changeflag{with increasing $\beta$, the CYM potential is approaching the perturbative HCL large-$\beta$ limit from below.}

We emphasize that, while previous studies showed the potential for different $\beta$ values as functions of $r/a$, we find that plotting it as a function of $m_D r$ makes the comparison more intuitive. Most importantly, this allows to \changeflag{study the classical potential even at large $\beta$ values while incorporating the dominant UV divergence into the mass.}%
\footnote{\prevchangeflag{By adjusting discretization parameters $a$ and $N$, we ensure that, despite the divergence of $m_{D}^{\text{HCL}} \sim 1/a^{1/2}$, relevant distance scales $r \sim 1/m_D$ lie within our lattice $r_{\text{min}} \ll 1/m_D \ll r_{\text{max}}$. Here the minimal and maximal lattice distances are given by $r_{\rm{min}} = a$ and $r_{\rm{max}} \sim Na$.}} 
We have performed such a large-$\beta$ extrapolation for the HCL calculations already in \se\ref{sec:HCL_results} by considering ${\rm Im}[V_{\rm cl}(r)]/g^2T$ as a function of $m_{D}^{\text{HCL}}\, r$ for increasing \changeflag{$\beta$.} 

Instead of the leading-order perturbative HCL mass $m_{D}^{\text{HCL}}$, one could instead plot the CYM results as a function of $\hat m_D\, r$, with the values of \changeflag{the screening mass} $\hat{m}_D$ extracted from CYM longitudinal chromo-electric field correlators as discussed in Sec.~\ref{sec:thermalstate}. The latter is an estimate of the Debye mass including higher-order and nonperturbative contributions. We find that, in practice, the HCL Debye mass \eqref{hclimv} is proportional to the correlator-extracted Debye mass $\hat{m}_D$ 
over a wide range of $\beta$ via \eq\eqref{eq:mdhat_md} with proportionality constant $1.42$, and that the resulting rescaling only introduces a constant rescaling of the horizontal axis. 
Since there are larger uncertainties inherent in the extraction of $\hat{m}_D$ from the chromo-electric correlators, we have chosen to rescale the classical lattice and HCL results using the same HCL mass definition.

\subsubsection{Small-distance behavior}

%%%%%%====FIGURE======%%%%%%%%%%%%%%%%%%%%%%%%%%%%%
\begin{figure}[t]
\centerline{
\includegraphics[width=0.48\linewidth]{\pToFigs/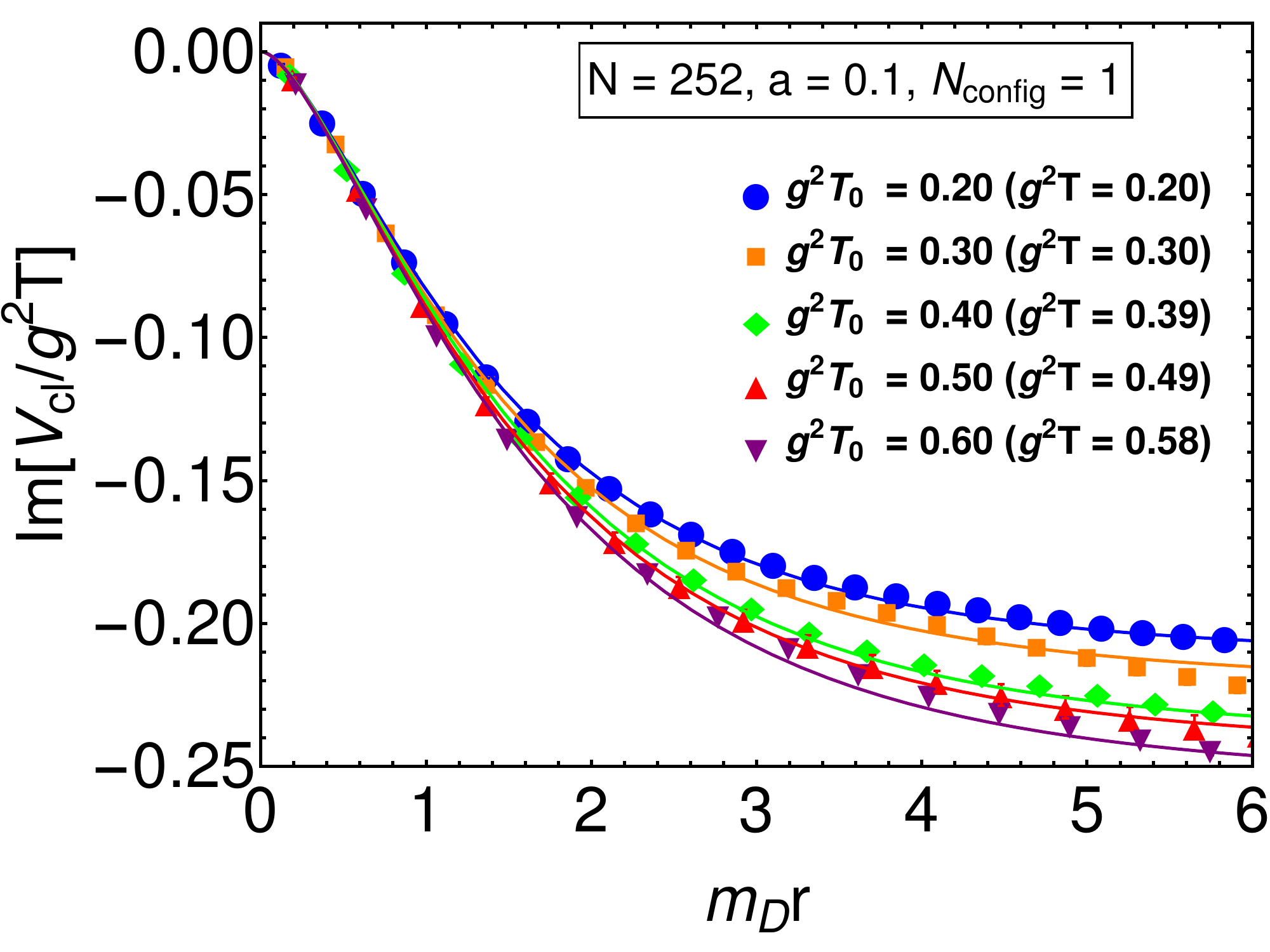}
$\;\;\;$
\includegraphics[width=0.48\linewidth]{\pToFigs/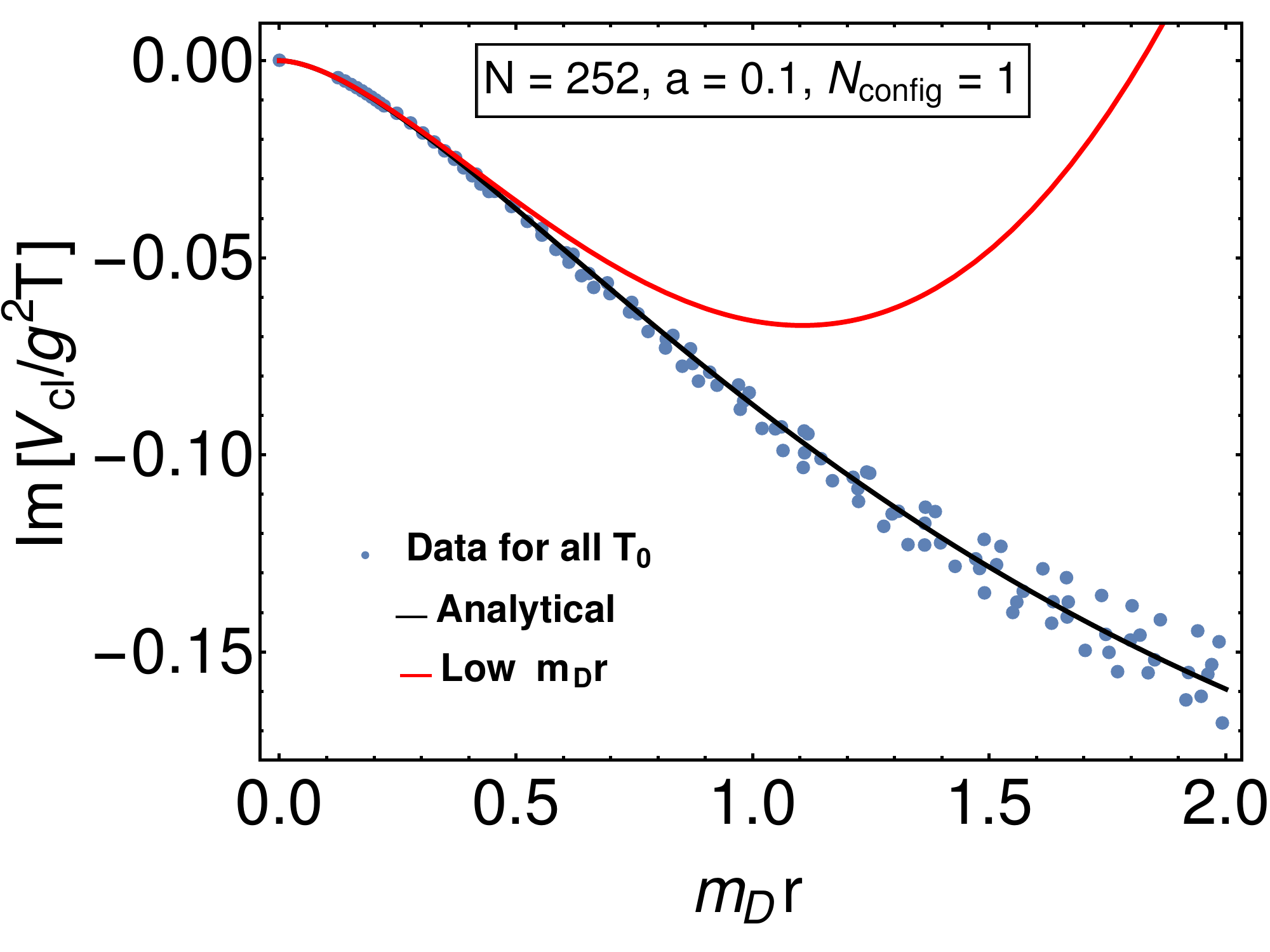}
}
\caption{
(Left) The extracted classical potential ${\rm Im}[V_{\rm cl}(r)]/g^2T$ as a function of $m_Dr$ for fixed lattice spacing $a = 0.1$ but for different temperatures $g^2 T = 0.2,\dots, 0.59$,
which correspond to the range $\beta = 300,\dots, 100$.
The continuous curves are fits to our data points using the form of the potential in \eqref{Aphi}.
(Right) The combined data of the left panel, with additional data sets for $g^2 T_0 = 0.25, 0.35, 0.45, 0.55, 0.65$, is compared to the analytical form \eqref{Aphi} with parameter values \eqref{eq:latt_AB_fit} and its low-distance limit \eqref{eq:ImV2_lowX} as black and red curves, respectively. 
}
\label{plot:short-dist}
\end{figure}
%%%%%%%%%%%%%%%%%%%%%%%%%%%%%%%%%%%%%%%%%%%%%%%%%%%%

\changeflag{Let us now compare the fitting function \eqref{Aphi} with parameters $A_\infty$ and $B$ to the imaginary part of the heavy-quark potential. 
We saw previously that it} agrees with this form in both the HTL and the HCL formalisms with the parameters
\begin{align}
\label{eq:HTL_AB_fit}
A^\text{HTL}_{\infty} &\approx -0.080\, C_F\,, \qquad \qquad \quad~ B^\text{HTL}=1 \\
 A_\infty^{\text{HCL}}(\beta {\to} \infty) &\approx -0.108\,C_F\,, \qquad B^{\text{HCL}}(\beta {\to} \infty) \approx 1.548\,.
 \label{eq:HCL_AB_fit}
\end{align}
We \changeflag{provide evidence} that, at small distances, the classical potential extracted from our lattice simulation data also follows the functional form \eqref{Aphi} \changeflag{for different values of $\beta$}. This is shown in the left panel of \fig\ref{plot:short-dist}, where ${\rm Im}[V_{\rm cl}(r)]/g^2T$ for fixed lattice spacing $a=0.1$ is plotted
for different temperatures corresponding to $100 \lesssim \beta \leq 300$, as a function of $m_D\, r$. Fits to each data set using \eqref{Aphi} in the considered interval $m_D r \leq 6$ are included as continuous lines. One finds very good agreement between the data and the respective fits. Moreover, although $m_D$ and the extracted potential vary with the temperature $T$ and, equivalently, with $\beta$, all curves are seen to fall on top of each other at low distances $m_D r \lesssim 1$. 
\changeflag{We have seen such behavior already in \fig\ref{plot:imv_mDr} for even smaller values of $\beta$.}%
\footnote{\changeflag{In \fig\ref{plot:imv_mDr}, curves for different $\beta$ values are observed to overlap at short distances $m_D r \lesssim 1$. This works best when the smallest distance on the lattice is much smaller than the inverse Debye mass. For $\beta = 16$, there is only one data point within $m_D r \lesssim 1$, making a comparison more challenging.}}
\changeflag{This indicates} approximately temperature-independent low-distance behavior for \changeflag{a wide} $\beta$ interval. 
Averaging over the resulting parameter values \changeflag{from the fits to each data set in \fig\ref{plot:short-dist}}, we obtain
\begin{align}
\label{eq:latt_AB_fit}
 A_\infty^{\text{latt}} = -0.24 \pm 0.02 = \left( -0.18 \pm 0.015 \right)C_F\,,
 \qquad
 B^{\text{latt}} = 1.2 \pm 0.05\,.
\end{align}
Here the errors have been estimated by adding the standard error and the fitting error. 

In the right panel of \fig\ref{plot:short-dist}, we combine all of the curves of the left panel and compare them to the function \eqref{Aphi} with the mean parameter values in \eqref{eq:latt_AB_fit}, shown as a black line. As expected, one observes very good agreement with our data, especially at low distances $m_D r$.
We can also perform a short-distance expansion of the fitting function for the imaginary part of the potential \eqref{Aphi}, neglecting terms $\mathcal{O}((m_D r)^4)$,
\begin{align}
 \label{eq:ImV2_lowX}
 {\rm Im}[V_{\rm cl}(r)] \simeq -r^2\,\frac{1}{9}\,|A_\infty|B^2\, g^2T\,m_D^2\left( 4-3\gamma - 3\log (B\, m_D\, r) \right).
\end{align}
\changeflag{The resulting curve} is shown as a red line and is observed to agree well with our data points for $m_D\, r \lesssim 0.5$. 
\changeflag{Thus, for a wide range of $\beta$ values, the short-distance behavior of our CYM lattice data agrees well with the perturbative functional form \eqref{Aphi} and its leading short-distance expansion, which is parametrically given by $\left|{\rm Im}[V_{\rm cl}(r)]\right| \sim C_F\, g^2T\,\left(m_D\,r\right)^2\,\log (m_D\,r)$.}

\section{Conclusions and outlook}
\label{sec:conclusions}

In this paper we used classical-statistical lattice simulations of pure Yang-Mills fields to extract the imaginary part of the heavy-quark potential as a function of the quark-antiquark separation.  In order to carry out our simulations on large lattices, we used a simplified scheme to generate thermalized gauge field configurations which relied on initialization of chromo-electric fields in momentum-space followed by a period of self-thermalization.  In order to determine whether the fields were thermalized, we extracted the temperature of the system as a function of time directly from the dynamically generated chromo-electric field correlators.

We went beyond previous classical-statistical lattice calculations of ${\rm Im}[V_{\rm cl}]$ by considering rather large lattice sizes and \changeflag{a wide range of values of $\beta$ (or, equivalently, of the lattice spacing $a$ and temperature)}.  We considered both $SU(2)$ and $SU(3)$ gauge groups and found that the potentials obtained obeyed a simple Casimir scaling. 
\changeflag{Using our CYM simulations on such large lattices, we argued that ${\rm Im}[V_{\rm cl}]$ should be plotted as a function of $m_D\, r$, with the screening mass $m_D$, when comparing our lattice results for different $\beta$ values or to perturbative calculations. In particular, we showed that ${\rm Im}[V_{\rm cl}]$ approaches the perturbative lattice-regularized HCL results with increasing $\beta$. As a side result, our calculations suggest that a finite large-$\beta$ limit exists for ${\rm Im}[V_{\rm cl}]$ when calculated in the HCL framework and plotted as a function of $m_D\, r$. In the region of small distances $m_D\, r \lesssim 1$, we demonstrated that ${\rm Im}[V_{\rm cl}]$ from our CYM lattice simulations is insensitive to the finite values of $\beta$ over a wide parameter range, and is close to the HCL results.}

We also found that both our lattice simulation and HCL results were very well approximated by a functional form which can be obtained from a leading-order hard-thermal loop calculation. One only needs to take into account a different \changeflag{prefactor $A_\infty$}
and a different scaling of the argument of the functional form, encoded in a parameter $B$.  Using fits of this form and then expanding the result at small $m_D r$, we were able to extract \changeflag{small-distance approximations for the imaginary part of the heavy-quark potential}.

Looking to the future, we plan to perform a similar calculation in an anisotropic gluonic plasma that is expanding along the longitudinal direction. This case is of phenomenological interest since the expansion of the quark-gluon plasma in relativistic heavy-ion collisions generates large momentum-space anisotropy. This renders the perturbative analytic calculation of the imaginary part of the heavy-quark potential ill-defined due to the presence of unstable modes in the soft gauge field propagator. \changeflag{In contrast, classical-statistical lattice simulations are applicable in that case, and this work sets the stage to corresponding future studies out of equilibrium}.

\begin{table*}[ht] 
{\small
\centering
 \begin{tabular}{|c c | c| c| c| c| c| c |} 
 \cline{3-8}
 \multicolumn{2}{c|}{} & \multicolumn{6}{c|}{${\rm Im}[V_{\rm cl}(r, t \rightarrow \infty)]/g^2T \qquad [a = 1]$} \\
 \hline
 $g^2 T_0$ & $g^2 T$ & $r/a = 1$ & $r/a = 2$ & $r/a = 3$ & $r/a = 4$ & $r/a = 5$ & $r/a = 6$   \\ [0.5ex] 
 \hline\hline
 0.20 & 0.185(2) & -0.030 & -0.081(1) & -0.131(2) & -0.174(2) & -0.210(3) & -0.240(4) \\ 
 \hline
 0.25 & 0.227(4) & -0.037(1) & -0.099(2) & -0.158(3) & -0.209(4) & -0.252(5) & -0.288(6) \\
 \hline
 0.30 & 0.267(5) & -0.045(1) & -0.117(2) & -0.187(4) & -0.246(5) & -0.297(6) & -0.341(8) \\
 \hline
 0.35 & 0.306(7) & -0.052(1) & -0.136(3) & -0.217(5) & -0.286(7) & -0.345(8) & -0.397(10) \\
 \hline
 0.40 & 0.343(11) & -0.060(2) & -0.157(5) & -0.249(8) & -0.329(11) & -0.398(13) & -0.461(15) \\  
 \hline
 0.45 & 0.379(11) & -0.069(2) & -0.179(5) & -0.284(8) & -0.376(11) & -0.457(14) & -0.529(16) \\  
 \hline
  0.50 & 0.413(11) & -0.078(2) & -0.204(5) & -0.323(9) & -0.428(12) & -0.515(14) & -0.574(19) \\ [1ex] 
 \hline 
\end{tabular}
}
\caption{Values of ${\rm Im}[V_{\rm cl}(r)]/g^2T$ for different $T_0$ extracted from lattice simulations with $a=1, N = 64$ and 224 configurations.}
\label{tab:table1}
\end{table*}

\begin{table*}[ht] 
{\small
\centering
 \begin{tabular}{|c c | c| c| c| c| c| c |} 
 \cline{3-8}
 \multicolumn{2}{c|}{} & \multicolumn{6}{c|}{${\rm Im}[V_{\rm cl}(r, t \rightarrow \infty)]/g^2T \qquad [a = 0.1]$} \\
 \hline
 $g^2 T_0$ & $g^2 T$ & $r/a = 10$ & $r/a = 20$ & $r/a = 30$ & $r/a = 40$ & $r/a = 50$ & $r/a = 60$   \\ [0.5ex] 
 \hline\hline
 0.20 & 0.199(1) & -0.104(1) & -0.165(1) & -0.188(2) & -0.200(2) & -0.207(3) & -0.210(3) \\ 
 \hline
 0.25 & 0.248(2) & -0.114(1) & -0.171(1) & -0.191(2) & -0.200(2) & -0.207(2) & -0.212(3) \\
 \hline
 0.30 & 0.296(2) & -0.128(1) & -0.186(2) & -0.207(2) & -0.224(2) & -0.232(3) & -0.236(4) \\
 \hline
 0.35 & 0.345(3) & -0.137(1) & -0.197(2) & -0.218(2) & -0.229(2) & -0.243(3) & -0.258(4) \\
 \hline
 0.40 & 0.394 & -0.146 & -0.206(1) & -0.226(1) & -0.239(2) & -0.255(3) & -0.269(4) \\  
 \hline
 0.45 & 0.442(3) & -0.152(1) & -0.209(2) & -0.225(2) & -0.235(3) & -0.248(3) & -0.255(4) \\  
 \hline
  0.50 & 0.490(8) & -0.161(3) & -0.218(4) & -0.238(4) & -0.242(5) & -0.251(5) & -0.257(10) \\ [1ex] 
 \hline 
\end{tabular}
}
\caption{Values of ${\rm Im}[V_{\rm cl}(r)]/g^2T$ for different $T_0$ extracted from lattice simulations with $a= 0.1, N = 252$ (one configuration).}
\label{tab:table2}
\end{table*}

\acknowledgments{
We would like to thank M.\ Escobedo and J.H.\ Weber for useful discussions.
M.S.\ and B.K.\ were supported by the U.S. Department of Energy, Office of Science, Office of Nuclear Physics Award No.~DE-SC0013470. We thank the Ohio Supercomputer Center for support under the auspices of Project No.\ PGS0253. The computational results presented have been also achieved in part using the Vienna Scientific Cluster (VSC).
}

\appendix

\setlength{\arrayrulewidth}{0.3mm}
\setlength{\tabcolsep}{7pt}
\renewcommand{\arraystretch}{1.5}

\section{Tables}
\label{app:tables}

Our simulation results of ${\rm Im}[V_{\rm cl}(r)]/g^2T$ for selected distances $r/a$ and different parameters $g^2T_0$ and the extracted temperatures $g^2 T$ are shown in Table \ref{tab:table1} for $a=1$ and in Table \ref{tab:table2} for $a=0.1$.

%%%%%%%%%%%%%%%%%%%%%%%%%%%%%%%%%%
%%%%%%%             BIBLIOGRAPHY	                  %%%%%%%
%%%%%%%%%%%%%%%%%%%%%%%%%%%%%%%%%%

%%%%%%%%%%%%%%%%%%%%%%%%%%%%%%%%%%%%%%%%%%%%%%%%%%%%%%%%%%%%%
\bibliographystyle{JHEP.bst}
\bibliography{ImV.bib}
%%%%%%%%%%%%%%%%%%%%%%%%%%%%%%%%%%%%%%%%%%%%%%%%%%%%%%%%%%%%%

\end{document}